\documentstyle[12pt,aaspp4,lscape,flushrt]{article}

\begin{document}

\title{The Redshift Evolution of the Metagalactic Ionizing Flux Inferred from Metal
Line Ratios in the Lyman Forest}
\author{Antoinette Songaila\altaffilmark{1}} 
\affil{Institute for Astronomy, University of Hawaii, 2680 Woodlawn Drive,
  Honolulu, HI 96822\\}

\altaffiltext{1}{The author was a visiting astronomer at the W. M. Keck
  Observatory, jointly operated by the California Institute of Technology and
  the University of California.}

\vskip 1in

\centerline{To be published in {\it Astronomical Journal,} 6/98}

\begin{abstract}

Metal line ratios in a sample of 13 quasar spectra obtained with the HIRES
spectrograph on the KeckI telescope have been analyzed to characterize the
evolution of the metagalactic ionzing flux near a redshift of 3.  The
evolution of \ion{Si}{4}/\ion{C}{4} has been determined using three different
techniques: using total column densities of absorption line complexes, as in
Songaila \& Cowie\markcite{sc96} (1996); using the column densities of
individual Voigt profile components within complexes; and using direct optical
depth ratios.  All three methods show that \ion{Si}{4}/\ion{C}{4} changes
abruptly at $z \sim 3$, requiring a jump in value of about a factor of 3.4,
and indicating a significant change in the ionizing spectrum that occurs
rapidly between $z = 2.9$\ and $z = 3$, just above the redshift at which
Reimers et al.\markcite{reim} (1997) detected patchy \ion{He}{2} ${\rm
Ly}\alpha$\ absorption.  At lower redshifts, the ionization balance is
consistent with a pure power law ionizing spectrum  but at higher redshifts the
spectrum must be very soft, with a large break at the He$^+$\ edge.  An
optical depth ratio technique is used to measure the abundances of ions whose
transitions lie within the forest and \ion{C}{3}, \ion{Si}{3} and
\ion{O}{6} are detected in this way.   The presence of a significant amount of
\ion{O}{6} at $z > 3$\ suggests either a considerable volume
of \ion{He}{3} bubbles embedded in the more general region where the ionizing
flux is heavily broken, or the addition of collisional ionization to the
simple photoionization models.

\end{abstract}


\section{Introduction} \label{intro}

Whereas we know from the absence of any significant Gunn-Peterson effect even
in the highest redshift quasars that hydrogen reionization of the
intergalactic gas must have taken place at $z > 5$, we have much less
information about the period at which the bulk of singly ionized helium
converted to doubly ionized helium.  Since late He$^+$\ ionization may
significantly change the temperature of the intergalactic gas it is critical
to understand this heating if we are to correctly model the growth of
structure in the IGM, and determine the mapping of the baryon density to
observable quantities such as observations of the neutral hydrogen ${\rm
Ly}\alpha$\ forest and the He$^+$\ ${\rm Ly}\alpha$\ opacity.
Phenomenological modelling of this event depends critically on the softness of
the composite spectrum of the ionizing sources (e.g.\ Miralda-Escud\'e \& 
Rees\markcite{mr93} 1993; Madau \& Meiksin\markcite{mm} 1994),
amplified by the subsequent radiative transfer, and such models cannot be
considered reliable in predicting the high energy ($E > 54~{\rm eV}$)
metagalactic ionizing spectrum above the He$^+$\ ionization edge, which
determines the fraction of singly ionized helium.

Our most direct information on the He$^+$\ opacity is through observations of
quasars whose spectra extend to the He$^+$\ ${\rm Ly}\alpha$\
wavelength. Despite the extreme difficulty of these measurements, successful
observations of the He$^+$\ ${\rm Ly}\alpha$\ absorption have been made toward
$z > 2.8$\ quasars with HST (Jakobsen et al.\markcite{jak} 1994; Hogan et
al.\markcite{hog} 1997; Reimers et al.\markcite{reim} 1997) and of the $z =
2.72$\ quasar HS1700+6416 with HUT (Davidsen et al.\markcite{dkz} 1996; Zheng
et al.\markcite{zdk} 1998).  The He$^+$\ ${\rm Ly}\alpha$\ opacity shows a
marked decrease from a value of $\tau = 3.2^{+\infty}_{-1.1}$\ in Q0302$-$003
at $z = 3.29$\ to $\tau = 1.0 \pm 0.07$\ in HS1700+6416.  More remarkably, the
Reimers et al.\markcite{reim} observation of the intermediate redshift quasar
HE2347$-$4342 at $z = 2.89$\ shows both `troughs' and `voids' in the
\ion{He}{2} ${\rm Ly}\alpha$\ observations, suggesting that at $z \sim 2.8$\ we are
seeing fully ionized \ion{He}{3} bubbles interspersed among as yet unionized
He$^+$\ regions and that it is at this point that the porosity of \ion{He}{3}
regions is approaching unity.

The recent discovery that the bulk of ${\rm Ly}\alpha$\ forest absorption with
$N({\rm H~I}) > 3\times 10^{14}~\rm cm^{-2}$\ contains associated metals
(Songaila \& Cowie\markcite{sc96} 1996, hereafter SC) gives us an
alternative approach to the problem since ionization balance in these forest
metals provides a diagnostic of the shape of the metagalactic flux in the
neighborhood of the He$^+$\ edge.  As was first noted in Songaila et
al.\markcite{shc} (1995), the value of \ion{Si}{4}/\ion{C}{4} in high
ionization systems is critically dependent on the He$^+$\ ionization edge
break strength.  This has subsequently been investigated in more detail by SC,
Savaglio et al.\markcite{sav} (1997) and Giroux \& Shull\markcite{gs97}
(1997), among others.  As will be discussed further here, the bulk of the
forest metal line systems at the currently observed redshifts ($z \sim 2 - 4$)
are high ionization (\ion{C}{2}/\ion{C}{4} $\ll 0.1$) and so, unless there is
a strong break at the edge, they will have
\ion{Si}{4}/\ion{C}{4} $\ll 0.1$\ even for the higher Si/C abundances
characteristic of low metallicity systems.  Therefore, once the He$^+$\ is
fully ionized and the IGM becomes relatively transparent to the integrated
quasar spectrum (e.g.\ in the metagalactic spectrum of Haardt \&
Madau\markcite{hm} 1996), the observed \ion{Si}{4}/\ion{C}{4} values should
fall in this low range.  SC and Savaglio et al.\markcite{sav} (1997) have
shown that, whereas this is generally true at $z < 3$, much higher
\ion{Si}{4}/\ion{C}{4} values are regularly seen at $z > 3$, suggesting a
significant change above this redshift, which would be consistent with the
interpretation of the Reimers et al.\markcite{reim} (1997) observations as
showing the redshift at which \ion{He}{3} bubbles begin to overlap. 
Boksenberg\markcite{bok} (1998) has recently questioned this result, based on
an analysis of the the redshift evolution of the ion ratios in the separate
Voigt profile components in complex systems rather than in the integrated
column densities of the complexes.  However, his analysis is based on rather a
small sample of systems without clear selection criteria.  In this paper, I
shall use the largest sample to date to demonstrate unambiguously that,
irrespective of the method of analysis, there is indeed a rapid jump in the
value of \ion{SI}{4}/\ion{C}{4} at a redshift just below 3, and that the
ionization stages in the metals are consistent with this being, for most
systems, the point at which they change from being ionized by a metalgalactic
spectrum that is, at $z > 3$, heavily broken above 54~eV to one that is only
mildly broken at the lower redshifts.

The sample and the data reduction are described in \S 2 and the reader who is
primarily interested in the results could safely skip this section.  The
evolution of the \ion{C}{2}/\ion{C}{4} and \ion{Si}{4}/\ion{C}{4} values with
redshift is described in \S 3 where I show that the presence of a jump in
\ion{Si}{4}/\ion{C}{4} values at $z$\ just under 3 is highly significant and
does not depend on the method of analysis, whether by total column densities
of complex, by the column densities of individual Voigt components, or by
directly analyzing the distribution of optical depth ratios.  In \S 4 I
consider the overall ionization balance including intermediate ionization
stages such as \ion{C}{3} and \ion{Si}{3} and high ions such as \ion{N}{5} and
\ion{O}{6} using, for those lines that lie primarily within the forest, the
optical depth distributions of the ensembles of such lines, which provides a
new and robust technique for determining their properties.  The overall
ionization balance is broadly consistent with unbroken power law
photoionization at $z < 3$; however, the observation of significant amounts of
\ion{O}{6} at $z > 3$\ requires either the presence of a considerable volume of
\ion{He}{3} bubbles permeating regions where the ionizing flux is heavily
broken, or the addition of collisional ionization to the simple photoionization
models.  Finally, the conclusions are briefly summarised in \S 5.

\section{Data} \label{data}

The sample was drawn from observations of 13 quasars with complete or
near-complete coverage between the quasars' Lyman alpha and \ion{C}{4}
emission lines.  All observations (summarised in Table~\ref{tbl:1}) were made
with the HIRES spectrograph on the KeckI 10m telescope, using a $1.14\arcsec
\times 7\arcsec$\ slit, which gives a resolution of $R = 37,000$, and with
total exposure times from 3.3 hours to just over 10 hours.  The approximate
S/N per resolution element for each quasar at wavelengths between ${\rm Ly}\alpha$\
and \ion{C}{4} emission is summarised in Table~\ref{tbl:1} and ranges from 60
(Q0256$-$000) to 330 (Q1422+231).

Individual exposures were generally set to 40 minutes (providing a reasonable
compromise between minimizing cosmic ray hits and read noise effects) and with
the 2048 Tektronix CCD read out with a binning of two in the spatial direction
and one in the spectral direction, resulting in a spatial pixel of
$0.38\arcsec$\ and $\Delta\lambda/\lambda = 7.3\times 10^{-6}$; this substantially
oversamples the resolution element.  For the higher redshift quasars several
settings of the HIRES cross-disperser were used to provide the full
wavelength coverage.  All but three of the quasars have full coverage of the
wavelength region of interest (Table~1, column 4) and the remaining three have
small interorder gaps.  In general, the spectra extend from approximately
$3800~{\rm \AA}$, below which the HIRES sensitivity begins to fall rapidly, up to
the wavelength of the quasar's \ion{C}{4} emission.  A white dwarf star and
the thorium-argon calibration lamp were also observed either immediately
before or immediately after each sequence of exposures on a quasar.

A relatively straightforward extraction technique was used since optimization
does not provide substantial gains for these spatially undersampled and
signal-dominated spectra.  Starting from approximate order and wavelength
solutions, the position and shape of each order on the CCD was determined from
the observations of the white dwarf.  Only orders lying fully on the CCD were
extracted.  A small number of known bad points and columns were cleaned by
interpolation, and a number of regions, such as the central blemish on the
chip, flagged as unusable.  The first pass cosmic ray rejection was made by
forming a residual image with a 5-point two-dimensional median subtracted and
then subtracting an 11-point row by row median.  Any points in this
differenced image that exceeded expected values by more than $4~\sigma$\ were
interpolated from neighboring points along the column (spatial) direction.

The spectral image was distorted by bilinear interpolation to make the echelle
order parallel to the rows. The sky was then determined from the median of
pixels on the same column lying between $2\arcsec$\ and $3.5\arcsec$\ on
either side of the maximum signal, and subtracted from the spectral image, and
the spectrum was extracted in a $\pm 1\arcsec$\ window.  The spectra for all the
sub-exposures were then combined with an optimal S/N weighting, and any
residual cosmic ray structure was removed by identifying any strongly deviant
($> 5~\sigma$) points in the individual spectrum and flagging and removing
these from a second and final summation.  The blaze function was determined
from a normalised fifth-order polynomial fit to the white dwarf spectrum,
iteratively repeated to exclude any absorption lines, and divided into the
spectrum.  The wavelength and spectrum for each order were then individually
stored.

The Th-Ar calibration spectrum was extracted in the same fashion as the object
spectrum but without any sky subtraction step.  For each order, each of the
stronger unblended calibration lines was fitted with a gaussian to determine
its centroid, and a third-order polynomial fit was then applied to determine
the wavelength solution.  From 10 to 30 calibration lines were used for each
order and the r.m.s. residual was normally less than $2\times 10^{-3}~{\rm \AA}$, so
that velocities should be accurate to better than $0.1~{\rm km\ s}^{-1}$.  Finally, the
wavelength scale was converted to vacuum heliocentric, which is used
throughout the present paper.

A typical extracted order longward of the forest is shown in
Figure~\ref{newfig:1a} which shows order 63 from a 2-hr exposure of
Q1159+123, obtained as three 40-minute exposures in 1997 April.  The
absorption lines in the center of the order are a \ion{C}{4} doublet, and the
rise in flux to shorter wavelengths reflects the onset of the quasar's
${\rm Ly}\alpha$\ and \ion{N}{5} emission.  Figure~\ref{newfig:1b} shows an order
in the forest blueward of the quasar's ${\rm Ly}\alpha$\ emission, where the accuracy
of the sky subtraction may be gauged from the precision with which the bases
of the saturated lines approach zero.

A final combined spectrum was made from all the observations taken at all
echelle settings, with spectra covering each order added in a S/N optimized
fashion to form a combined spectrum of that order.  The final 8-hr spectrum of
order 63 in Q1159+123 is shown in Figure~\ref{newfig:2} for comparison with
Figure~\ref{newfig:1a}.   An associated effective exposure time vector,
normalised by the S/N, was also maintained at each point;  this can be
combined with the counts to determine the noise at each point in the spectrum.
For some purposes it is useful to have a combined spectrum and this was formed
by interpolating all the orders to a uniform ($0.05~{\rm \AA}$) grid which provides
Nyquist sampling throughout the wavelength range.  A typical spectrum (of
Q1159+123) is shown as Figure~\ref{newfig:3}.  This spectrum has a Lyman limit
system just longward of $4150~{\rm \AA}$\ and stretches through to \ion{C}{4}
emission just below $7000~{\rm \AA}$, near the atmospheric A band, which can be
clearly seen in the spectrum.  Most of the remaining features between
${\rm Ly}\alpha$\ and \ion{C}{4} emission are \ion{C}{4} absorption line doublets.

For most purposes normalized (continuum fitted) spectra from each order were
used.  The continuum fitting is made on a $\pm 2000~{\rm km\ s}^{-1}$\ region of interest
by iterating a second-order polynomial fit to a 10-point smoothed spectrum and
rejecting significantly deviant regions ($\vert\Delta({\rm spectrum})/{\rm
fit}\vert > 0.025$).  The fit to the region around a \ion{C}{4} complex at
$z = 3.26101$\ in Q1159+123 is shown in Figure~\ref{newfig:4}.  Multiple Voigt
profiles were then fitted to the absorption line profiles using the IDL
version of Bevington's CURVEFIT algorithm.

\section{Line Ratio Evolution with Redshift} \label{ratio}

\subsection{Introduction}

Each quasar spectrum was first searched for all complexes of \ion{C}{4}
doublets redward of the L$\alpha$\ forest and outside of the quasar's
proximity zone, defined to stretch to $-4000~{\rm km\ s}^{-1}$\ blueward of the quasar's
redshift.  The sample was then restricted to those cases for which \ion{SI}{4}
also lies redward of the quasar's ${\rm Ly}\alpha$\ emission.  The redshift range
used for each quasar is summarised in Table~\ref{tbl:1}.  Voigt profiles were
next fitted to each such complex.  The choice of the number of Voigt
components is somewhat arbitrary because of the problems of blending but the
complexes are quite weak in general and in the absence of saturation the
integrated column densities of complexes should be well determined.  In the
same way, integrated column densities or upper limits for \ion{Si}{4},
\ion{C}{2} and \ion{Si}{2} were determined for lines lying longward of the
forest.  The usable redshift range for \ion{C}{2} is also shown in
Table~\ref{tbl:1}. The profile fits to the \ion{C}{4} and \ion{Si}{4} doublets
in all the cloud complexes with $10^{13}~{\rm cm}^{-2} < N({\rm C~IV}) <
10^{14}~{\rm cm}^{-2}$ are shown in Figure~5.  The redshifts, number
of Voigt components (a minimum of two was used), \ion{C}{4} column density,
\ion{Si}{4} column density, and \ion{C}{2} and \ion{Si}{2} column densities
(where these lines are redward of the forest) are summarised for all complexes
with $N({\rm C~IV}) > 5 \times 10^{12}~{\rm cm}^{-2}$\ in Table~\ref{tbl:2}.

\subsection{Evolution of total column density ratios}

Figure~\ref{fig:2} shows the evolution of the integrated value of 
\ion{Si}{4}/\ion{C}{4} in the 76 complexes with $5 \times 10^{12}~\rm cm^{-2} \le
N({\rm C~IV}) \le 10^{14}~\rm cm^{-2}$\ as well as in the the higher redshift
systems satisfying this condition in Savaglio et al.'s\markcite{sav} (1996)
observations of Q0000$-$263.  As SC and Savaglio et al.\markcite{sav} (1996)
previously noted, \ion{Si}{4}/\ion{C}{4} increases considerably between
$z = 2$\ and $z = 4$, with a strong suggestion of a break in the distribution
of values around $z = 3$.  In contrast, the \ion{C}{2}/\ion{C}{4} ratio shown
in Figure~\ref{fig:3} does not exhibit this behaviour: in most of the
complexes where \ion{C}{2}/\ion{C}{4} can be measured, it is weak (${\rm
C~II/C~IV} < 0.1$) or undetected, regardless of redshift.  Median values of
\ion{Si}{4}/\ion{C}{4} are shown as a function of redshift in
Figure~\ref{fig:4}.  The median ratio is $0.039^{+0.009}_{-0.006}$\ for all
the systems below $z = 3$\ but $0.13^{+0.04}_{-0.04}$\ above $z = 3$, where
the errors are $1~\sigma$\ computed using the median sign method.  A key
diagnostic in interpreting the data is whether the evolution is in the form of
a step function or is a smoother change with redshift.  A discontinuous change
might arise from the effects of a change in the metagalactic ionizing spectrum
at the He$^+$\ ionization edge, whereas a smoother evolution might follow from
the more general evolution of the external ionizing field and the range of
internal densities and internal ionization in the complexes.  The data of
Figure~\ref{fig:2} and Figure~\ref{fig:4} are strongly suggestive of an abrupt
transition between $z = 2.9$\ and $z = 3$.  This redshift is slightly higher
than, but comparable to, the redshift ($ z = 2.8$\ to 2.9) below which
\ion{He}{2} would be inferred to be predominantly ionized to \ion{He}{3} from
the direct observations of the \ion{He}{2} ${\rm Ly}\alpha$\ forest (Reimers
et al.\markcite{reim} 1997).

To quantify this, distribution functions of \ion{Si}{4}/\ion{C}{4} have been
formed for complexes with $z = 2.75 - 2.95$, just below the suspected break,
and for $z = 3.0 - 3.2$, just above it.  The histograms of
\ion{Si}{4}/\ion{C}{4} values for the 16 systems in the low redshift interval
and the 22 systems in the high redshift interval are shown in
Figure~\ref{fig:5}.  The median value of $0.038^{+0.029}_{-0.010}$\ at $2.75 <
z < 2.95$\ is consistent with that of all the $z < 3$\ systems, whereas the
$3.0 < z < 3.2$\ median value of $0.14^{+0.04}_{-0.075}$\ is consistent with
that of all the $z > 3$\ systems.  Moreover, a rank sum test shows that the $z
= 2.75 - 2.95$\ data are consistent with being drawn from the same
distribution as the $z < 2.75$\ systems, and the $z = 3.0 - 3.2$\ data with
being drawn from the same distribution as the $z > 3.2$\ systems.  However,
the rank sum test shows that the $z = 3.0 - 3.2$\ data has a probability of
only $2.2 \times 10^{-4}$\ of being drawn from the same distribution as the $z
= 2.75 - 2.95$\ data, and that at the 98\% confidence level the data require
a jump of at least a multiplicative factor of 1.8 from the lower to the higher
redshift interval.  The data  are therefore consistent with invariant
distributions at high and low redshift together with an abrupt jump between $z
= 2.9$\ and $z = 3$.  The corresponding step function, with
\ion{Si}{4}/\ion{C}{4} changing by a factor of 3.4 at $z = 2.95$, 
is shown as the dashed line in Figure~\ref{fig:4}.

Assuming instead that the evolution of the median
ionization ratio is a smooth power law in $(1+z)$\ requires an evolution of
the form $(1+z)^{5.4}$\ which is shown as the dotted line in
Figure~\ref{fig:4}.  This however predicts a jump of only 1.57 between $z =
2.85$\ and $z = 3.1$\ which, as already noted, is not large enough to
reconcile the relative distribution functions of Figure~\ref{fig:5}.  It would
therefore seem that simple smooth power law evolution is not consistent with
the data.  

\subsection{Internal structure}

Variations of the ionization ratios in different regions within a complex can
result from structure in the internal density of material in the complex, or
from the presence of internal ionization sources.  It is therefore of
considerable interest to examine the ionization ratios within the complexes to
understand how these effects can contribute to the overall ionization balance
and to determine whether or not this can change the conclusions of section
3.2. 

The simplest approach to this is to break down the comlexes into the
components determined by Voigt profile fitting (Boksenberg\markcite{bok}
1998).  Figure~\ref{fig:6} shows the redshift evolution of the
\ion{Si}{4}/\ion{C}{4} components broken down in this way.  All components
with $N({\rm C~IV}) > 5 \times 10^{12}~\rm cm^{-2}$\ have been included
irrespective of the total column density of the complex, so that a small
number of high density complexes provide a large number of components, and
these also include a number of Lyman limit systems in which radiative transfer
effects are significant.  However, even with these effects and with the
increased confusion owing to the large spread in components, the change
between high and low redshifts is clearly seen.  Figure~\ref{fig:7} shows
histograms of the distributions above and below $z = 3$.  If the components
are considered to be statisticaly independent samples then the median ratio is
$0.043^{+0.015}_{-0.008}$\ below $z = 3$\ and $0.15^{+0.04}_{-0.04}$\ above $z
= 3$.  At the 98\% confidence level, the minimum multiplicative increase to
bring the distributions into consistency is 1.93.

However, the methodology of fitting Voigt profiles is a very unsatisfactory
approach to this problem because of the considerable arbitrariness in the
choice of components, which can have several unfortunate effects, including
forcing other ions preferentially into or out of a particular component,
blending components, and introducing non-independent components.  These
problems can be avoided by using the more direct technique of computing the
ratios of the optical depths of various ions throughout the complex.  This
procedure is illustrated in Figures~\ref{fig:8} and \ref{fig:9}.  For each
identified \ion{C}{4} complex, all data points lying above some limiting
optical depth (here $\tau({\rm C~IV~1550}) > 0.05$) were identified, and
optical depths in the corresponding regions of the other ions were then
measured.  \ion{Si}{4} and \ion{C}{4} optical depths for a complex at $z =
3.33714$\ in 2000$-$330 are shown in Figure~\ref{fig:8}.  The ratio ($\tau
\lambda f$) relative to that of \ion{C}{4} then gives the ratio of $N({\rm
ion})/N({\rm C~IV})$\ throughout the velocity range of the complex, as is
illustrated in Figure~\ref{fig:9}, where it is compared with the average ion
ratios in the complex, shown as the solid line.  This procedure can be
slightly in error if there is a substantial amount of thermal broadening but
should be adequate for these heavier elements where the broadening is
primarily kinematic.
\ion{Si}{4}/\ion{C}{4} values determined in this way for all the complexes of
Table~\ref{tbl:2} is shown in Figure~\ref{fig:10} where again a clear
evolution of the distribution can be seen.

In order to determine the noise levels associated with this procedure, a large
number of blank field realizations were constructed, in which the redshift was
slightly displaced from the true value.  This Monte Carlo technique is extremely
powerful, providing an unbiased measure of the noise, including the systematic
effects of continuum uncertainty, varying S/N among the quasars and within a
single quasar spectrum, and contamination by unidentified absorption lines.
Indeed, this last-named property allows a robust measure (\S 4) of the average
column densities of ions lying in the Lyman alpha forest.

The measured distribution (solid line) of \ion{Si}{4}/\ion{C}{4} is compared
with the average of the random realizations (dashed line) in
Figure~15 for all the systems, and also for those above and below $z
= 3$.
\ion{Si}{4}/\ion{C}{4} is clearly seen to be much stronger above $z = 3$.
Figure~\ref{fig:12} shows median \ion{Si}{4}/\ion{C}{4} values as a function of
redshift, with errors derived from the distribution of medians in the
blank-field realizations.  The data again require a rapid jump at
around $z = 3$, with values of $0.035 \pm 0.005$\ below $z = 3$\ and $0.10 \pm
0.01$\ above.

\section{Ionization Ratios} \label{ion}

\subsection{Introduction}

The local shape of the ionizing flux and the value of the ionization parameter
may be estimated from the relative distribution of the ionization
levels of the various species.  Most work to date has concentrated on using
\ion{Si}{4}/\ion{C}{4} versus \ion{C}{2}/\ion{C}{4}, which, as we have
discussed in the introduction, is a powerful discriminant of the ionization
parameter and of the strength of any break near the He$^+$\ edge.  This
diagnostic also relies on absorption lines that can be measured outside the
forest.  The \ion{Si}{4}/\ion{C}{4} versus \ion{C}{2}/\ion{C}{4} method is
reconsidered (\S~4.2) in the light of the present larger data set.  However,
the optical depth ratio method developed in \S 3.3 provides a powerful way to
measure the average abundances of a number of other species (\ion{Si}{3},
\ion{C}{3}, \ion{N}{5}, and \ion{O}{6}) that lie primarily within the 
${\rm Ly}\alpha$\ forest.  We use this technique in \S 4.3 to determine the ensemble
properties of these ions as a function of redshift and to show that they can
be fitted into a consistent interpretation of the evolution of the shape of
the metagalactic flux.

\subsection{Si~IV/C~IV versus C~II/C~IV}

\ion{Si}{4}/\ion{C}{4} versus \ion{C}{2}/\ion{C}{4} is shown in
Figure~\ref{fig:13} for the 29 complexes with $10^{13}~\rm cm^{-2} < N({\rm
C~IV}) < 10^{14}~\rm cm^{-2}$\ and for which the \ion{C}{2} line lies longward
of the ${\rm Ly}\alpha$\ forest.  Open squares denote systems with $z > 3$\
and solid squares those with $z < 3$, while downward and/or leftward pointing
arrows correspond to systems in which \ion{Si}{4} or \ion{C}{2} is not
detected, with the point positioned at the $1~\sigma$\ level.  The curves show
a range of models computed with the CLOUDY code (Ferland\markcite{fer} 1993)
for a power-law ionizing spectrum with a range of breaks at the He$^+$\ edge
at 4~Rydberg, and with $\log_{10}({\rm Si/C}) = -0.66$, typical of metal-poor
systems.  As was found previously in SC, while many of the low redshift
systems are consistent with photoionization by a pure power law, most of the
high redshift systems have \ion{Si}{4}/\ion{C}{4} values which are much too
high relative to \ion{C}{2}/\ion{C}{4} for this to be the case.  They are
consistent with photoionization by heavily broken power laws with breaks
across the He$^+$\ edge in excess of 20 and softness ratios, $S \equiv
f_{\nu}(4\ {\rm Ryd})/f_{\nu}(1\ {\rm Ryd})$\ in excess of 300.

The models of SC were computed with simple broken power laws and a more
extensive discussion of more general spectra may be found in Giroux \& 
Shull\markcite{gr97} (1997).  However, Figure~\ref{fig:14} illustrates that the 
general behaviour is fairly model invariant; it compares the data with a wide
range of power laws ($-1.5$\ to $-2.5$) again computed with $\log_{10}({\rm
Si/C}) = -0.66$.  Unless Si/C is made unreasonably high at $z > 3$, not even
the softest power law spectrum (the top curve) can approach the high
\ion{Si}{4}/\ion{C}{4} values.  The introduction of the break at 4~Ryd moves
the curves to a new asymptote, as is illustrated by the dashed and dotted
curves.  Both are computed with an ionization parameter of $-1.6$\ and a $-2$\
power law spectrum below 4~Ryd, but with a range of break strengths.  The
dotted curve corresponds to the same models as in Figure~\ref{fig:13} in which
the spectrum above the break is extrapolated with the same spectral index as
it has below the break but at the lower flux level, as might be the case if
there was a substantial galaxy contribution to the ionizing spectrum.  The
dashed line is computed instead with a model in which the break is a He$^+$\
opacity whose cross section reduces as $E^{-2.7}$\ at higher energies.  This
latter model asymptotes to the same \ion{Si}{4}/\ion{C}{4} values but requires
a larger softness ratio of greater than about 2000 to achieve this.  It is
this effect that gave rise to Rauch et al.'s\markcite{rhs97} (1997) incorrect
conclusion that He$^+$\ breaks could not account for the high values of
\ion{Si}{4}/\ion{C}{4} in high ionization systems since they used a recovering
spectrum and an intermediate softness parameter (a flux reduction of 100 at
4~Ryd).  However, as the break becomes large, both models reach the same
limiting values which are consistent with the observed high values of
\ion{Si}{4}/\ion{C}{4} and low values of \ion{C}{2}/\ion{C}{4}.

In this asymptotic limit of a large break at the He$^+$\ edge, the
\ion{Si}{4}/\ion{C}{4} to \ion{C}{2}/\ion{C}{4} ratio approaches a single
track 
characterized essentially
uniquely by the ionization parameter, which determines, modulo a very weak
dependence on spectral index, the \ion{C}{2}/\ion{C}{4} value.  Therefore most
of both the high and low redshift systems are consistent with a relatively
small range of ionization parameter, $\log_{10}\Gamma = -1$\ to $-2$, but the
higher redshift systems mostly require the high energy spectrum to be blocked
above 54~eV whereas at the low redshifts this is not generally the case.
It is perhaps important to emphasise that this effect is not observationally
subtle and in Figure~19 we show \ion{C}{4}, \ion{Si}{4} and
\ion{C}{2} profiles for a sample of the high \ion{Si}{4}/\ion{C}{4} systems
over the redshift range.  For the high \ion{Si}{4}/\ion{C}{4} systems at low
redshift, \ion{C}{2} is generally clearly present but many of the high-$z$\ 
high \ion{Si}{4}/\ion{C}{4} systems are simply missing \ion{C}{2} absorption.

\subsection{Other ion ratios}

The optical depth technique discussed in section 3 can also be used to determine
the average properties of other important ions such as \ion{C}{3}, \ion{Si}{3}
and \ion{O}{6} whose transitions lie below the wavelength of ${\rm Ly}\alpha$\ and
for which individual line strengths cannot be reliably determined because of
forest contamination.

As with \ion{Si}{4}, the value of $\tau({\rm X})\lambda_XF_X/\tau({\rm
C~IV})\lambda_{\rm C~IV}f_{\rm C~IV}$\ was determined for the ion, $X$, at all
positions at which significant \ion{C}{4} absorption was present (again, $\tau({\rm
C~IV}\, 1550) > 0.05$) and a distribution of this quantity was constructed.  A
large number of random realizations were then made in which the \ion{C}{4}
redshift was slightly displaced, to determine the backgrounds and errors.
Because of forest contamination, which acts as an effective noise source, the
errors are larger than for species that lie outside the forest, such as
\ion{Si}{4}.  

Figure~20 shows the distributions of \ion{C}{3}/\ion{C}{4},
\ion{N}{5}/\ion{C}{4} and \ion{O}{6}/\ion{C}{4} for all the available systems,
and separately for $z>3$\ and $z<3$\ systems where the division is possible.
The median and mean values and their $1~\sigma$\ errors for a range of ions
are given in Tables~\ref{tbl:3} and \ref{tbl:4} respectively.  As is expected,
\ion{O}{6} is strongly detected, with a median $N({\rm O~VI})/N({\rm C~IV}) =
1.97 \pm 0.28$.  \ion{C}{3}, \ion{Si}{3} and \ion{N}{5} are more weakly
detected, with a median $N({\rm C~III})/N({\rm C~IV}) = 0.75 \pm 0.25$.
Neither \ion{Si}{2} nor \ion{C}{2} is significantly detected.

The ionization patterns above and below $z = 3$\ are summarised in
Figure~21 which uses the more robust median values.  The points show
the quantities $A_C\,N({\rm X})/A_X\,N({\rm C~IV})$, where $(A_X/A_C)$\ is the
assumed abundance of the element $X$\ with respect to C, which is taken to be
$[{\rm O/C}] = 0.5$, $[{\rm Si/C}] = 0.4$ and $[{\rm N/C]} = -0.7$,
characteristic of low metallicity systems. This quantity is plotted against
the ionization potential of the ion.

The primary difference between $z < 3$\ and $z > 3$\ remains the much higher
abundance of \ion{Si}{4} at the higher redshift.  The fractions of the high
ions such as \ion{N}{5} and \ion{O}{6} are marginally larger at $z < 3$\ than
at $z > 3$, but the effect is not large.  The $z < 3$\ ionization distribution
can be reproduced quite well by a very simple model in which the ionizing
spectrum is a $-1.8$\ power law and $\Gamma = -1.6$;  this is shown as the solid
bars in Figure~\ref{fig:17a}.  This highly oversimplified model is of course
only a rough representation of the ensemble average of conditions in the IGM.
However, for $z > 3$\ we need a mix of conditions.  As is discussed above,
most of the systems must lie in the regions exposed to highly broken spectra
in order to reproduce the lower ion ratios, particularly
\ion{Si}{4}/\ion{C}{4}, but some fraction of the volume must still be ionized
by spectra containing higher energy photons to reproduce the observed
\ion{O}{6}, if this is assumed to be produced by photoionization.  A model in
which 70\% of the systems are irradiated by a fully broken spectrum with
$\alpha = -1.8$\ below the break and $\Gamma = -1.5$, and 30\% of the volume
is occupied by a pure $\alpha = -1.8$\ spectrum with $\Gamma = -1.0$\ provides
a reasonable representation, as is show by the bars in Figure~\ref{fig:17b}.

\section{Conclusions} \label{conc}

I summarise the results of the paper by noting that, irrespective of the
analysis methodology adopted, there is a significant change in the ionization
balance of forest metal lines which occurs just below a redshift of 3.  At
lower redshifts, the ionization balance in the forest lines is fully
consistent with a pure power law ionization spectrum with an index of $-1.8$\
but at higher redshifts the high values of \ion{Si}{4}/\ion{C}{4} seen in most
of the forest clouds despite generally low \ion{C}{2}/\ion{C}{4} values
implies that the ionizing flux must be very soft, with a large break at the
He$^+$\ edge.  The change occurs quite rapidly between $z = 2.9$\ and $z = 3$,
just above the redshift at which highly patchy \ion{He}{2} ${\rm Ly}\alpha$\
absorption is seen in the quasar HE~2347$-$4342 (Reimers et
al.\markcite{reim} 1997).  The simplest explanation seems to be that we are
seeing the redshift at which \ion{He}{2} ionizes completely to \ion{He}{3} as
the \ion{He}{3} Str\"omgren spheres overlap.

\acknowledgments
I am grateful to the many people at the Keck telescopes who assisted with
these observations and most particularly to Steve Vogt whose HIRES
spectrograph made them possible.  I would also like to thank Len Cowie and
Esther Hu for obtaining some of the observations on which this work is based,
Dieter Reimers for computing positions and magnitudes for unpublished bright
quasars, and Sandra Savaglio for providing updated information on higher
redshift \ion{Si}{4}/\ion{C}{4} values.  The research was supported by the
National Science Foundation under grant AST 96-17216.

\newpage

\begin{deluxetable}{lccccccc}
\tablewidth{500pt}
\tablecaption{Observations \label{tbl:1}}
\tablehead{
\colhead{Quasar} & \colhead{Mag.} & \colhead{Expo.} &  \colhead{S/N} 
& \colhead{Complete} 
& \colhead{$z_{low}$} &
 \colhead{$z_{low}$}  & \colhead{$z_{high}$} 
\\ [0.5ex]  
&& (hrs) &&&  (Si~IV) & (C~II) 
}
\startdata
0014+813     & 16.5 & 7.65  & 120 & Yes  & 2.81  & 3.00 & 3.31 \nl
0256$-$000   & 18.2 & 5.33  & 60 & Yes  & 2.78   & 2.97 & 3.28 \nl
0302$-$003   & 17.8 & 10.67 & 90 & Yes  & 2.72   & 2.91 & 3.21 \nl
0636+680     & 16.5 & 6.08  & 130 & Yes  & 2.62  & 2.80 & 3.10 \nl
0741+4741    & 16.1 & 4.0   & 145 & No   & 2.66  & 2.84 & 3.15 \nl
0956+122     & 17.5 & 6.67  & 90 & Yes  & 2.74   & 2.92 & 3.23 \nl
1159+123     & 17.5 & 8.0   & 85 & Yes  & 2.90   & 3.09 & 3.42 \nl
1422+231     & 16.5 & 9.3   & 330 & Yes  & 3.01  & 3.21 & 3.54 \nl
1623+269     & 16.0 & 3.33  & 60 & Yes  & 2.12   & 2.27 & 2.48 \nl
1700+6416    & 16.1 & 8.33  & 260 & Yes  & 2.27  & 2.43 & 2.68 \nl
2000$-$330   & 19.0 & 4.67  & 60 & Near & 3.14   & 3.36 & 3.69 \nl
2126$-$158   & 17.3 & 4.45  & 130 & No   & 2.72  & 2.90 & 3.21 \nl
2347$-$4342  & 16.3 & 6.67  & 130 & Yes  & 2.37  & 2.54 & 2.82

\enddata
\end{deluxetable}
\clearpage

\newpage

\begin{deluxetable}{llcrrrrr}
\tablewidth{500pt}
\tablecaption{C~IV, Si~IV, C~II and Si~II Column Densities \label{tbl:2}}
\tablehead{
\colhead{Quasar} & \colhead{$z$}   & 
\colhead{\# cpts.} 
& \colhead{$N({\rm C~IV})$} & \colhead{$N({\rm Si~IV})$} 
& \colhead{$N({\rm C~II})$} & \colhead{$N({\rm Si~II})$} 
}
\startdata

    1623+269  &  2.16141  &   3  &  $ 6.1 \times 10^{13}$  &  $ 1.4 \times 10^{12}$  &  \nodata \ \ \ \ \   &  $-1.9 \times 10^{11}$  \nl
    1623+269  &  2.24450  &   2  &  $ 3.1 \times 10^{13}$  &  $-3.2 \times 10^{10}$  &  \nodata \ \ \ \ \   &  $-2.4 \times 10^{11}$  \nl
   1700+6416  &  2.28952  &   2  &  $ 6.8 \times 10^{12}$  &  $ 6.2 \times 10^{10}$  &  \nodata \ \ \ \ \   &  $-2.6 \times 10^{11}$  \nl
   1700+6416  &  2.30826  &   3  &  $ 3.4 \times 10^{13}$  &  $ 1.7 \times 10^{12}$  &  \nodata \ \ \ \ \   &  $-9.5 \times 10^{11}$  \nl
   1700+6416  &  2.31536  &  13  &  $ 1.9 \times 10^{15}$  &  $ 6.1 \times 10^{13}$  &  \nodata \ \ \ \ \   &  $ 9.5 \times 10^{12}$  \nl
   1700+6416  &  2.37984  &   2  &  $ 1.3 \times 10^{13}$  &  $ 7.3 \times 10^{11}$  &  \nodata \ \ \ \ \   &  $ 4.6 \times 10^{11}$  \nl
    1623+269  &  2.40085  &   5  &  $ 1.2 \times 10^{14}$  &  $-1.8 \times 10^{11}$  &  $ 9.7 \times 10^{10}$  &  $ 4.5 \times 10^{11}$  \nl
   1700+6416  &  2.43304  &   7  &  $ 4.3 \times 10^{13}$  &  $ 9.0 \times 10^{12}$  &  \nodata \ \ \ \ \   &  $-1.2 \times 10^{11}$  \nl
   2347-4342  &  2.43814  &   2  &  $ 7.5 \times 10^{12}$  &  $ 3.0 \times 10^{11}$  &  \nodata \ \ \ \ \   &  $-3.4 \times 10^{10}$  \nl
   1700+6416  &  2.43863  &   8  &  $ 5.2 \times 10^{13}$  &  $ 2.0 \times 10^{12}$  &  $ 2.6 \times 10^{11}$  &  $-1.3 \times 10^{12}$  \nl
    1623+269  &  2.44353  &   3  &  $ 1.7 \times 10^{13}$  &  $ 6.8 \times 10^{10}$  &  $-3.1 \times 10^{12}$  &  $-1.2 \times 10^{12}$  \nl
    1623+269  &  2.44534  &   2  &  $ 6.3 \times 10^{12}$  &  $ 1.0 \times 10^{12}$  &  $ 2.7 \times 10^{12}$  &  $ 1.4 \times 10^{10}$  \nl
   1700+6416  &  2.56817  &   2  &  $ 6.6 \times 10^{12}$  &  $ 2.6 \times 10^{11}$  &  $-5.5 \times 10^{10}$  &  $ 2.0 \times 10^{10}$  \nl
   1700+6416  &  2.57857  &   4  &  $ 3.6 \times 10^{13}$  &  $ 1.3 \times 10^{12}$  &  $ 1.9 \times 10^{12}$  &  $-3.6 \times 10^{11}$  \nl
   2347-4342  &  2.63449  &   2  &  $ 9.9 \times 10^{12}$  &  $ 1.8 \times 10^{10}$  &  $ 9.9 \times 10^{12}$  &  $-3.0 \times 10^{11}$  \nl
   0741+4741  &  2.67283  &   2  &  $ 1.0 \times 10^{13}$  &  $ 8.7 \times 10^{10}$  &  \nodata \ \ \ \ \   &  $ 1.3 \times 10^{11}$  \nl
    0636+680  &  2.68228  &   2  &  $ 8.4 \times 10^{12}$  &  $ 7.8 \times 10^{11}$  &  \nodata \ \ \ \ \   &  $ 1.0 \times 10^{12}$  \nl
   0741+4741  &  2.69235  &   2  &  $ 6.5 \times 10^{13}$  &  $ 4.0 \times 10^{12}$  &  \nodata \ \ \ \ \   &  $-2.7 \times 10^{11}$  \nl
   0741+4741  &  2.69478  &   2  &  $ 2.7 \times 10^{13}$  &  $ 6.8 \times 10^{11}$  &  \nodata \ \ \ \ \   &  $-5.5 \times 10^{11}$  \nl
   0741+4741  &  2.71446  &   2  &  $ 6.3 \times 10^{12}$  &  $ 2.8 \times 10^{11}$  &  \nodata \ \ \ \ \   &  $ 1.9 \times 10^{11}$  \nl
    2126-158  &  2.72796  &   3  &  $ 2.8 \times 10^{13}$  &  $ 2.3 \times 10^{12}$  &  \nodata \ \ \ \ \   &  $-6.1 \times 10^{11}$  \nl
   0741+4741  &  2.73323  &   2  &  $ 5.1 \times 10^{13}$  &  $ 5.3 \times 10^{12}$  &  \nodata \ \ \ \ \   &  $-1.1 \times 10^{12}$  \nl
   2347-4342  &  2.73564  &   5  &  $ 1.1 \times 10^{14}$  &  $ 1.9 \times 10^{13}$  &  $ 6.8 \times 10^{12}$  &  $ 4.8 \times 10^{10}$  \nl
    2126-158  &  2.76916  &  12  &  $ 4.6 \times 10^{14}$  &  $ 6.4 \times 10^{13}$  &  \nodata \ \ \ \ \   &  $ 6.5 \times 10^{14}$  \nl
   0741+4741  &  2.78488  &   2  &  $ 1.0 \times 10^{13}$  &  $ 3.9 \times 10^{11}$  &  \nodata \ \ \ \ \   &  $-1.6 \times 10^{11}$  \nl
    0302-003  &  2.78663  &   9  &  $ 7.1 \times 10^{13}$  &  $ 3.1 \times 10^{12}$  &  \nodata \ \ \ \ \   &  $ 4.9 \times 10^{11}$  \nl
    2126-158  &  2.81959  &   5  &  $ 3.0 \times 10^{13}$  &  $ 2.0 \times 10^{12}$  &  \nodata \ \ \ \ \   &  $ 1.7 \times 10^{11}$  \nl
    0256-000  &  2.82444  &   6  &  $ 2.3 \times 10^{13}$  &  $ 1.2 \times 10^{12}$  &  \nodata \ \ \ \ \   &  $-7.1 \times 10^{11}$  \nl
    0302-003  &  2.82593  &   2  &  $ 6.8 \times 10^{12}$  &  $-2.4 \times 10^{11}$  &  \nodata \ \ \ \ \   &  $ 2.1 \times 10^{11}$  \nl
    0956+122  &  2.83185  &   2  &  $ 1.3 \times 10^{13}$  &  $ 8.3 \times 10^{10}$  &  \nodata \ \ \ \ \   &  $-6.9 \times 10^{11}$  \nl
    0956+122  &  2.83433  &   2  &  $ 8.9 \times 10^{12}$  &  $ 2.3 \times 10^{11}$  &  \nodata \ \ \ \ \   &  $-2.1 \times 10^{11}$  \nl
    0256-000  &  2.83599  &   2  &  $ 5.6 \times 10^{12}$  &  $ 4.2 \times 10^{11}$  &  \nodata \ \ \ \ \   &  $ 8.0 \times 10^{11}$  \nl
    0636+680  &  2.86883  &   2  &  $ 7.3 \times 10^{12}$  &  $ 5.0 \times 10^{11}$  &  $-3.9 \times 10^{11}$  &  $ 4.7 \times 10^{11}$  \nl
    0956+122  &  2.88806  &   4  &  $ 1.3 \times 10^{13}$  &  $ 5.1 \times 10^{11}$  &  \nodata \ \ \ \ \   &  $ 1.1 \times 10^{12}$  \nl
    0636+680  &  2.89163  &   5  &  $ 2.8 \times 10^{13}$  &  $ 5.6 \times 10^{10}$  &  $ 4.5 \times 10^9$  &  $-6.6 \times 10^{11}$  \nl
    0636+680  &  2.90374  &   9  &  $ 3.0 \times 10^{14}$  &  $ 1.4 \times 10^{14}$  &  $ 4.3 \times 10^{14}$  &  $ 5.5 \times 10^{13}$  \nl
   0741+4741  &  2.90455  &   6  &  $ 2.4 \times 10^{13}$  &  $ 7.8 \times 10^{11}$  &  $ 2.2 \times 10^{12}$  &  $-1.4 \times 10^{12}$  \nl
    2126-158  &  2.90708  &   2  &  $ 3.6 \times 10^{13}$  &  $ 8.7 \times 10^{12}$  &  $ 5.5 \times 10^{12}$  &  $-3.1 \times 10^{11}$  \nl
    0014+813  &  2.90823  &   2  &  $ 6.8 \times 10^{12}$  &  $ 4.5 \times 10^{11}$  &  \nodata \ \ \ \ \   &  $ 6.5 \times 10^{11}$  \nl
    0956+122  &  2.91490  &   2  &  $ 8.7 \times 10^{12}$  &  $ 2.1 \times 10^{11}$  &  \nodata \ \ \ \ \   &  $-3.4 \times 10^8$  \nl
    0302-003  &  2.91749  &   2  &  $ 1.8 \times 10^{13}$  &  $ 5.0 \times 10^{11}$  &  $ 1.3 \times 10^{12}$  &  $ 4.0 \times 10^{11}$  \nl
    0302-003  &  2.95882  &   2  &  $ 7.2 \times 10^{12}$  &  $ 1.1 \times 10^{11}$  &  $-1.1 \times 10^{12}$  &  $-3.5 \times 10^{11}$  \nl
    2126-158  &  2.96346  &   3  &  $ 2.0 \times 10^{13}$  &  $ 5.5 \times 10^{12}$  &  $ 1.6 \times 10^{11}$  &  $ 9.8 \times 10^{11}$  \nl
   0741+4741  &  2.96530  &   3  &  $ 1.0 \times 10^{13}$  &  $-1.3 \times 10^{11}$  &  $ 2.6 \times 10^{11}$  &  $ 6.7 \times 10^{11}$  \nl
    2126-158  &  2.96747  &   5  &  $ 2.5 \times 10^{13}$  &  $ 3.8 \times 10^{12}$  &  $ 9.1 \times 10^{12}$  &  $ 7.0 \times 10^{11}$  \nl
    0302-003  &  2.99540  &   2  &  $ 6.6 \times 10^{12}$  &  $ 3.2 \times 10^{11}$  &  $-1.9 \times 10^{12}$  &  $-2.4 \times 10^{11}$  \nl
    0302-003  &  3.00295  &   2  &  $ 3.3 \times 10^{13}$  &  $ 1.7 \times 10^{12}$  &  $ 7.4 \times 10^{11}$  &  $ 1.2 \times 10^{12}$  \nl
    0956+122  &  3.01043  &   2  &  $ 9.7 \times 10^{12}$  &  $ 1.5 \times 10^{12}$  &  $ 3.8 \times 10^{12}$  &  $-1.0 \times 10^{12}$  \nl
    0636+680  &  3.01303  &   2  &  $ 1.3 \times 10^{13}$  &  $ 5.8 \times 10^{11}$  &  $-1.4 \times 10^{12}$  &  $-7.8 \times 10^{11}$  \nl
    0636+680  &  3.01749  &   2  &  $ 4.1 \times 10^{13}$  &  $ 1.2 \times 10^{12}$  &  $ 7.0 \times 10^{11}$  &  $-4.1 \times 10^9$  \nl
   0741+4741  &  3.01763  &   6  &  $ 6.6 \times 10^{13}$  &  $ 2.5 \times 10^{13}$  &  $ 1.1 \times 10^{15}$  &  $ 1.5 \times 10^{14}$  \nl
    0256-000  &  3.01799  &   2  &  $ 1.5 \times 10^{13}$  &  $ 1.7 \times 10^{12}$  &  $ 4.3 \times 10^{10}$  &  $-1.2 \times 10^{12}$  \nl
   0741+4741  &  3.03469  &   3  &  $ 1.0 \times 10^{13}$  &  $ 9.7 \times 10^{11}$  &  $ 5.9 \times 10^{10}$  &  $-2.3 \times 10^{11}$  \nl
    0302-003  &  3.04705  &   2  &  $ 5.7 \times 10^{12}$  &  $-8.7 \times 10^{10}$  &  $ 3.2 \times 10^{11}$  &  $-5.7 \times 10^{10}$  \nl
    0956+122  &  3.05287  &   2  &  $ 1.0 \times 10^{13}$  &  $ 1.8 \times 10^{12}$  &  $-3.3 \times 10^{11}$  &  $ 6.1 \times 10^{10}$  \nl
   0741+4741  &  3.05366  &   4  &  $ 1.1 \times 10^{13}$  &  $ 6.3 \times 10^{12}$  &  $-1.7 \times 10^{12}$  &  $ 1.7 \times 10^{11}$  \nl
    0256-000  &  3.08434  &   2  &  $ 4.5 \times 10^{13}$  &  $ 1.3 \times 10^{13}$  &  $ 6.6 \times 10^{12}$  &  $ 1.7 \times 10^{11}$  \nl
    1422+231  &  3.09030  &   4  &  $ 4.2 \times 10^{13}$  &  $ 3.1 \times 10^{12}$  &  \nodata \ \ \ \ \   &  $-1.3 \times 10^{11}$  \nl
    0956+122  &  3.09691  &   2  &  $ 7.1 \times 10^{12}$  &  $ 1.0 \times 10^{12}$  &  $ 1.5 \times 10^{12}$  &  $ 8.6 \times 10^{11}$  \nl
    0956+122  &  3.11418  &   2  &  $ 6.9 \times 10^{13}$  &  $ 3.9 \times 10^{12}$  &  $ 1.2 \times 10^{12}$  &  $-5.2 \times 10^{11}$  \nl
    1422+231  &  3.13413  &   3  &  $ 1.6 \times 10^{13}$  &  $ 9.7 \times 10^{11}$  &  \nodata \ \ \ \ \   &  $-3.3 \times 10^{10}$  \nl
    1422+231  &  3.13702  &   2  &  $ 6.2 \times 10^{12}$  &  $ 1.1 \times 10^{12}$  &  \nodata \ \ \ \ \   &  $ 5.3 \times 10^{11}$  \nl
    0956+122  &  3.15294  &   5  &  $ 3.9 \times 10^{13}$  &  $ 1.2 \times 10^{13}$  &  $ 4.2 \times 10^{12}$  &  $ 3.3 \times 10^{11}$  \nl
    1159+123  &  3.16701  &   2  &  $ 8.6 \times 10^{12}$  &  $ 5.5 \times 10^{11}$  &  $-3.6 \times 10^{11}$  &  $ 1.5 \times 10^{11}$  \nl
    2000-330  &  3.17249  &   2  &  $ 1.3 \times 10^{13}$  &  $ 6.7 \times 10^{12}$  &  \nodata \ \ \ \ \   &  $ 2.1 \times 10^{12}$  \nl
    0956+122  &  3.17860  &   3  &  $ 1.6 \times 10^{13}$  &  $ 2.3 \times 10^{12}$  &  $ 8.8 \times 10^{11}$  &  $-3.0 \times 10^{11}$  \nl
    2000-330  &  3.19155  &   5  &  $ 7.6 \times 10^{13}$  &  $ 1.9 \times 10^{13}$  &  \nodata \ \ \ \ \   &  $ 6.2 \times 10^{13}$  \nl
    0256-000  &  3.19849  &   2  &  $ 1.6 \times 10^{13}$  &  $ 2.4 \times 10^{11}$  &  $ 6.3 \times 10^{11}$  &  $ 2.6 \times 10^{11}$  \nl
    0956+122  &  3.22310  &   5  &  $ 7.5 \times 10^{13}$  &  $ 1.5 \times 10^{13}$  &  $ 2.0 \times 10^{12}$  &  $ 1.9 \times 10^{11}$  \nl
    1159+123  &  3.22554  &   2  &  $ 2.6 \times 10^{13}$  &  $ 2.2 \times 10^{12}$  &  $-1.3 \times 10^{12}$  &  $-2.7 \times 10^{11}$  \nl
    0014+813  &  3.22623  &   5  &  $ 5.8 \times 10^{13}$  &  $ 3.0 \times 10^{12}$  &  $-1.6 \times 10^{12}$  &  $-1.3 \times 10^{12}$  \nl
    0014+813  &  3.23339  &   2  &  $ 9.6 \times 10^{12}$  &  $ 1.0 \times 10^{12}$  &  $ 2.0 \times 10^{11}$  &  $ 7.0 \times 10^{10}$  \nl
    1159+123  &  3.26101  &   2  &  $ 3.8 \times 10^{13}$  &  $ 2.6 \times 10^{12}$  &  $-4.0 \times 10^{11}$  &  $ 7.6 \times 10^{11}$  \nl
    2000-330  &  3.33281  &   4  &  $ 6.2 \times 10^{13}$  &  $ 2.9 \times 10^{13}$  &  \nodata \ \ \ \ \   &  $ 3.8 \times 10^{12}$  \nl
    2000-330  &  3.33714  &   2  &  $ 2.7 \times 10^{13}$  &  $ 3.7 \times 10^{12}$  &  \nodata \ \ \ \ \   &  $ 8.2 \times 10^{11}$  \nl
    1159+123  &  3.37844  &   2  &  $ 1.6 \times 10^{13}$  &  $ 2.2 \times 10^{12}$  &  $ 1.3 \times 10^{11}$  &  $-5.2 \times 10^9$  \nl
    1422+231  &  3.38277  &   5  &  $ 4.4 \times 10^{13}$  &  $ 2.8 \times 10^{12}$  &  $ 2.1 \times 10^{12}$  &  $ 3.7 \times 10^{11}$  \nl
    1159+123  &  3.40448  &   2  &  $ 1.9 \times 10^{13}$  &  $ 2.3 \times 10^{12}$  &  $-6.8 \times 10^{11}$  &  $ 3.5 \times 10^{11}$  \nl
    1422+231  &  3.41121  &   2  &  $ 8.7 \times 10^{12}$  &  $ 1.7 \times 10^{12}$  &  $ 4.7 \times 10^{10}$  &  $ 7.9 \times 10^{10}$  \nl
    1422+231  &  3.44717  &   2  &  $ 4.0 \times 10^{13}$  &  $ 7.6 \times 10^{12}$  &  $-4.0 \times 10^{10}$  &  $ 3.9 \times 10^{10}$  \nl
    2000-330  &  3.50532  &   2  &  $ 9.7 \times 10^{12}$  &  $ 4.1 \times 10^{12}$  &  $ 2.6 \times 10^{12}$  &  $ 2.3 \times 10^{10}$  \nl
    1422+231  &  3.53863  &  11  &  $ 1.8 \times 10^{14}$  &  $ 3.8 \times 10^{13}$  &  $ 4.4 \times 10^{13}$  &  $ 2.4 \times 10^{12}$  \nl
    2000-330  &  3.54937  &  11  &  $ 2.3 \times 10^{14}$  &  $ 6.3 \times 10^{13}$  &  $ 1.7 \times 10^{14}$  &  $ 9.4 \times 10^{12}$ 

\enddata
\end{deluxetable}
\clearpage

\newpage

\begin{deluxetable}{lccc}
\tablewidth{400pt}
\tablecaption{Medians of Column Density Ratios With Respect to C~IV \label{tbl:3}}
\tablehead{
\colhead{Ion} & \colhead{All} & \colhead{$z > 3$} &  \colhead{$z < 3$} 
}
\startdata
C~II      & $0.02\pm0.03$    & $0.01\pm0.04$    & $0.03\pm0.09$     \nl
C~III     & \nodata          & $0.75\pm0.25$    & \nodata           \nl
C~IV      & 1                &  1               & 1                 \nl
Si~II     & $-0.001\pm0.004$ & $0.004\pm0.010$  & $-0.004\pm0.006$  \nl
Si~III    & $0.029\pm0.010$  & $0.08\pm0.037$   & $0.013\pm0.019$   \nl
Si~IV     & $0.056\pm0.003$  & $0.10\pm0.005$   & $0.035\pm0.004$   \nl
N~V       & $0.019\pm0.008$  & $0.018\pm0.024$  & $0.019\pm0.006$   \nl
O~VI      & $1.97\pm0.28$    & $1.59\pm0.21$    & $2.75\pm0.78$

\enddata
\end{deluxetable}
\clearpage

\newpage

\begin{deluxetable}{lccc}
\tablewidth{400pt}
\tablecaption{Means of Column Density Ratios With Respect to C~IV \label{tbl:4}}
\tablehead{
\colhead{Ion} & \colhead{All} & \colhead{$z > 3$} &  \colhead{$z < 3$} 
}
\startdata
C~II      & $-0.28$     & $-0.03$    & $-0.47$     \nl
C~III     & \nodata     & $1.97$     & \nodata     \nl
C~IV      & 1           &  1         & 1           \nl
Si~II     & $0.08$      & $0.27$     & $-0.07$     \nl
Si~III    & $0.0$       & $0.07$     & $-0.04$     \nl
Si~IV     & $0.12$      & $0.17$     & $0.07$      \nl
N~V       & $0.05$      & $0.09$     & $0.02$      \nl
O~VI      & $1.8$       & $1.3$      & $2.2$

\enddata
\end{deluxetable}
\clearpage

\begin{landscape}

\begin{figure}
\figurenum{1a}
\plotfiddle{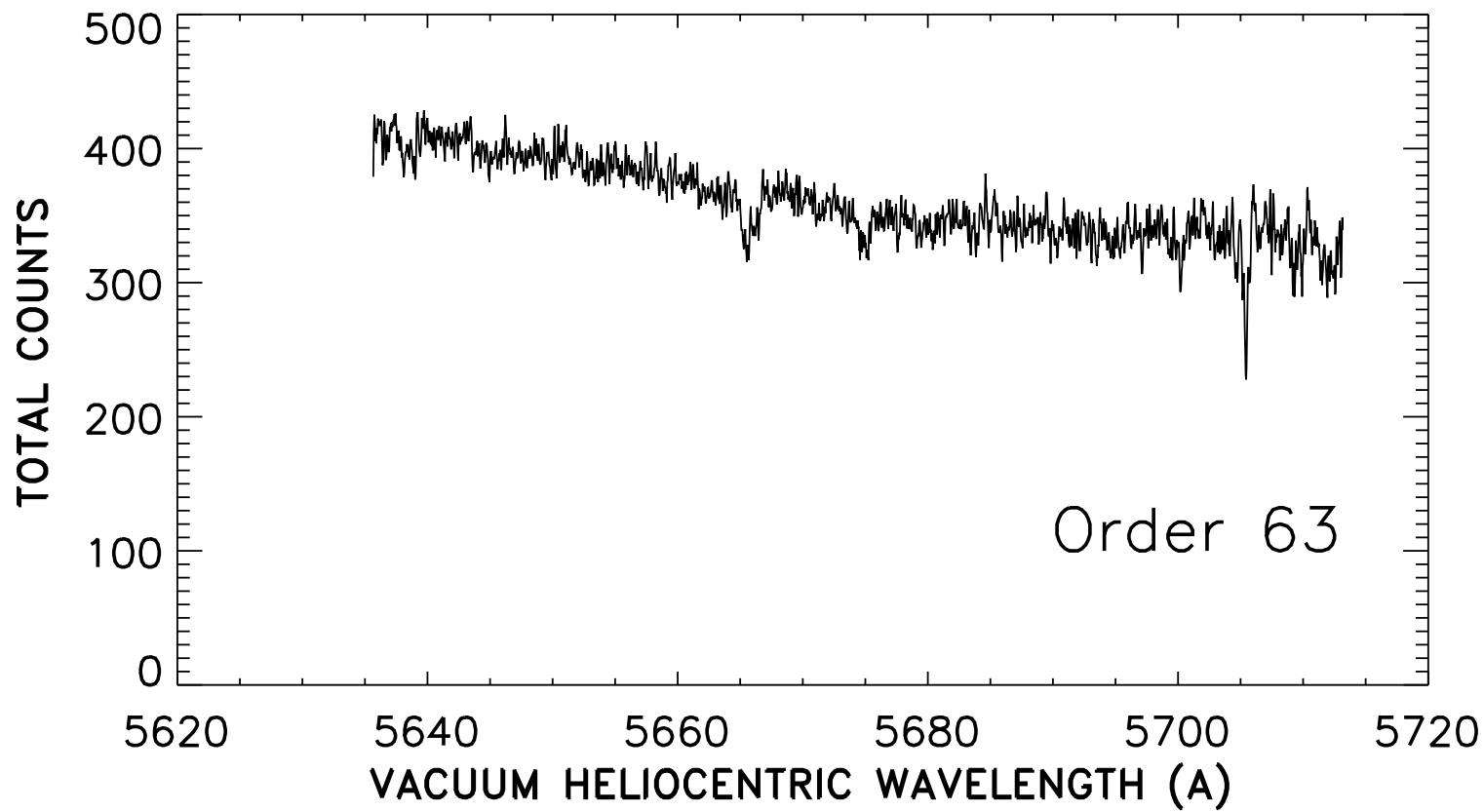}{5.5in}{0.0}{120}{120}{-300}{0}
\caption{A sample extraction of one of the orders
longward of the Ly$\alpha$\ forest in the quasar Q1159+123.  The total exposure
is 2~hr, consisting of three 40~m integrations.  The spectrum is unsmoothed
and oversamples the resolution element.}\label{newfig:1a}
\end{figure}

\begin{figure}
\figurenum{1b}
\plotfiddle{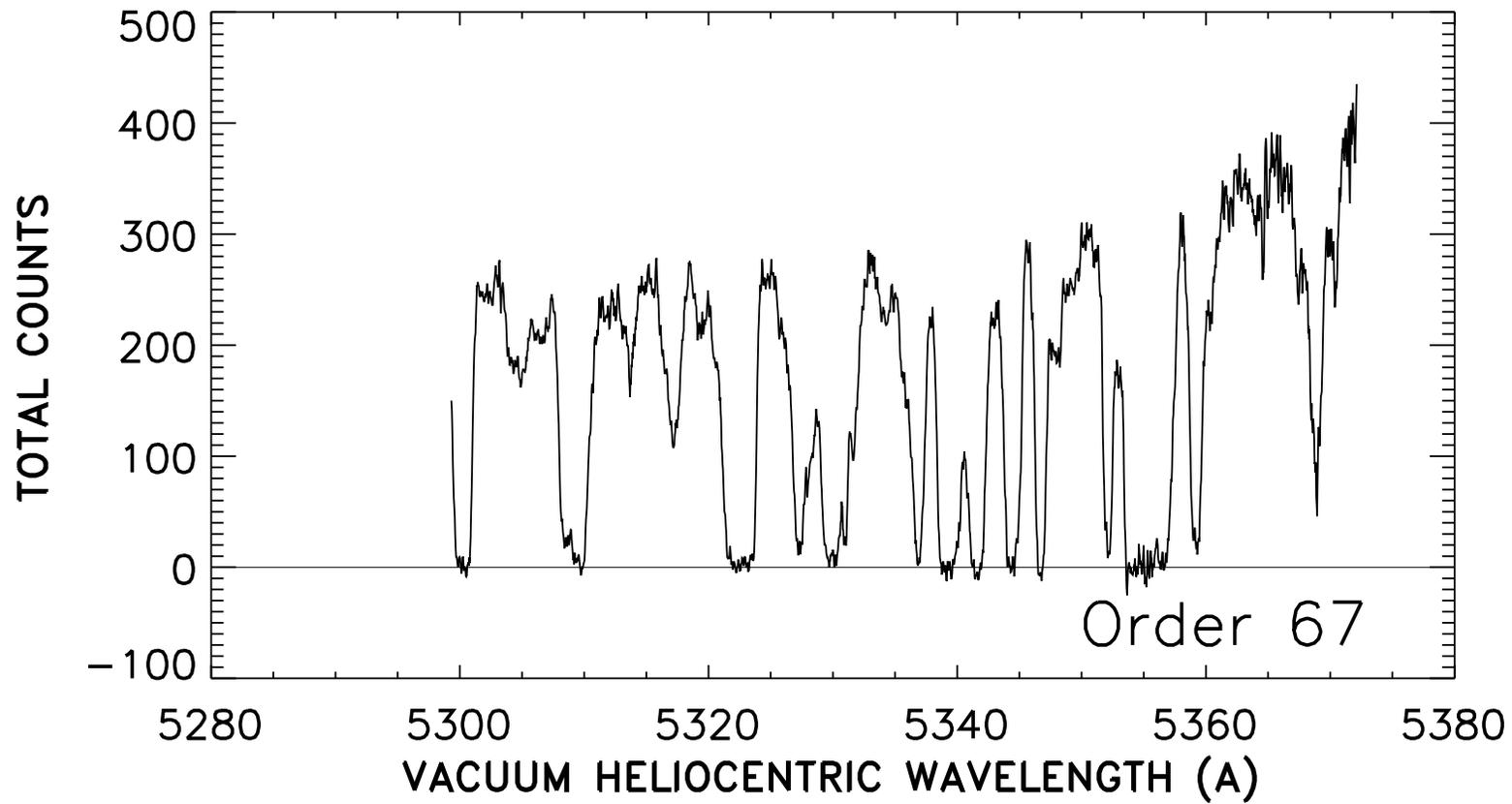}{5.5in}{0.0}{120}{120}{-300}{0}
\caption{An order in the same spectrum
as (a) but lying in the Ly$\alpha$\ forest.  The precision of the sky
subtraction may be assessed from the degree to which the bases of the
saturated Ly$\alpha$\ lines approximate to zero.  
}\label{newfig:1b}
\end{figure}

\begin{figure}
\figurenum{2}
\plotfiddle{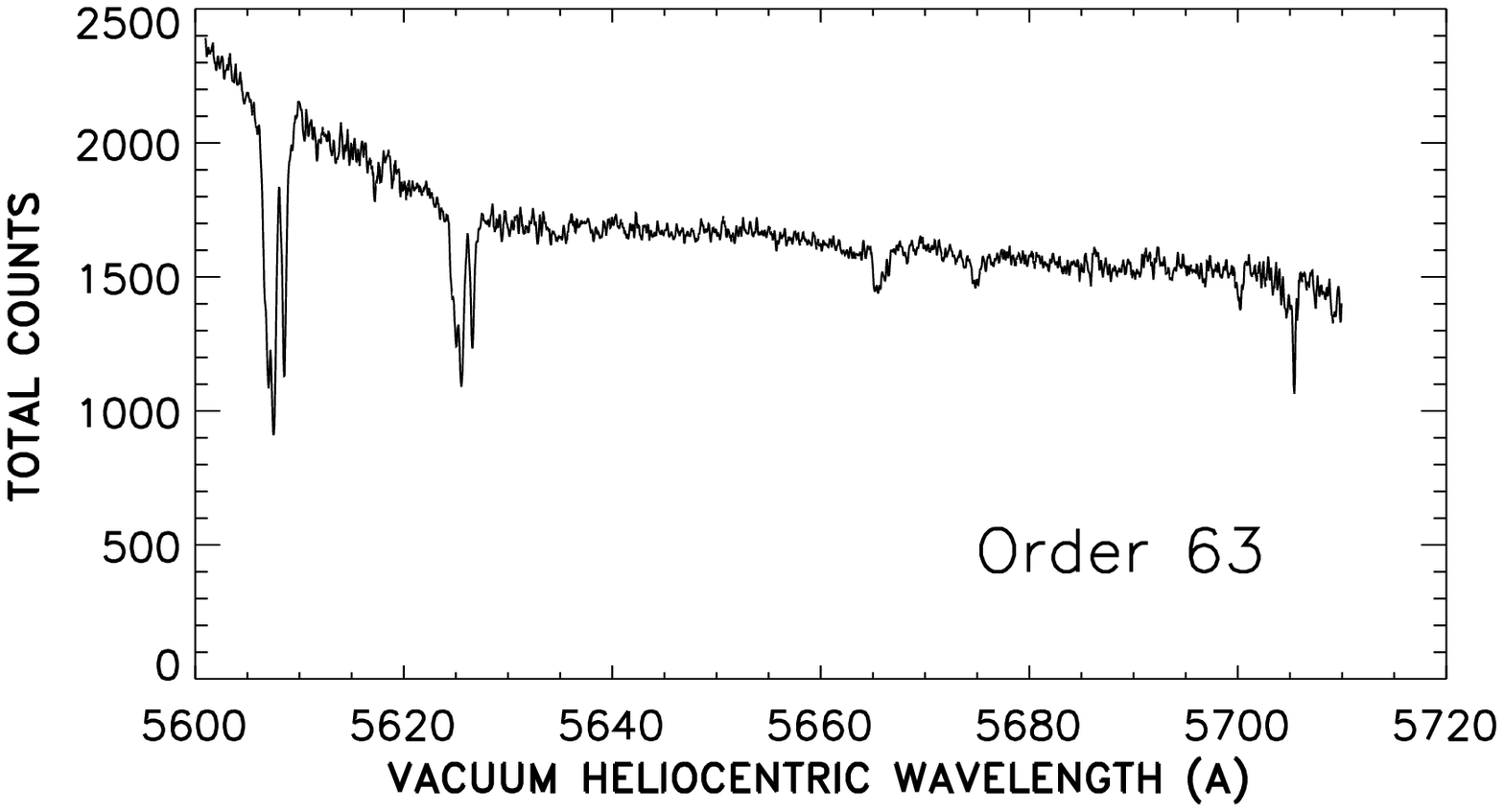}{5.5in}{0.0}{120}{120}{-300}{0}
\caption{Final processed and extracted order in Q1159+123
for comparison with Figure~1(a).  The total exposure of 8~hr is combined from a
number of sub-exposures with spectra covering each order added in a S/N
optimized fashion.
}\label{newfig:2}
\end{figure}

\begin{figure}
\figurenum{3}
\plotfiddle{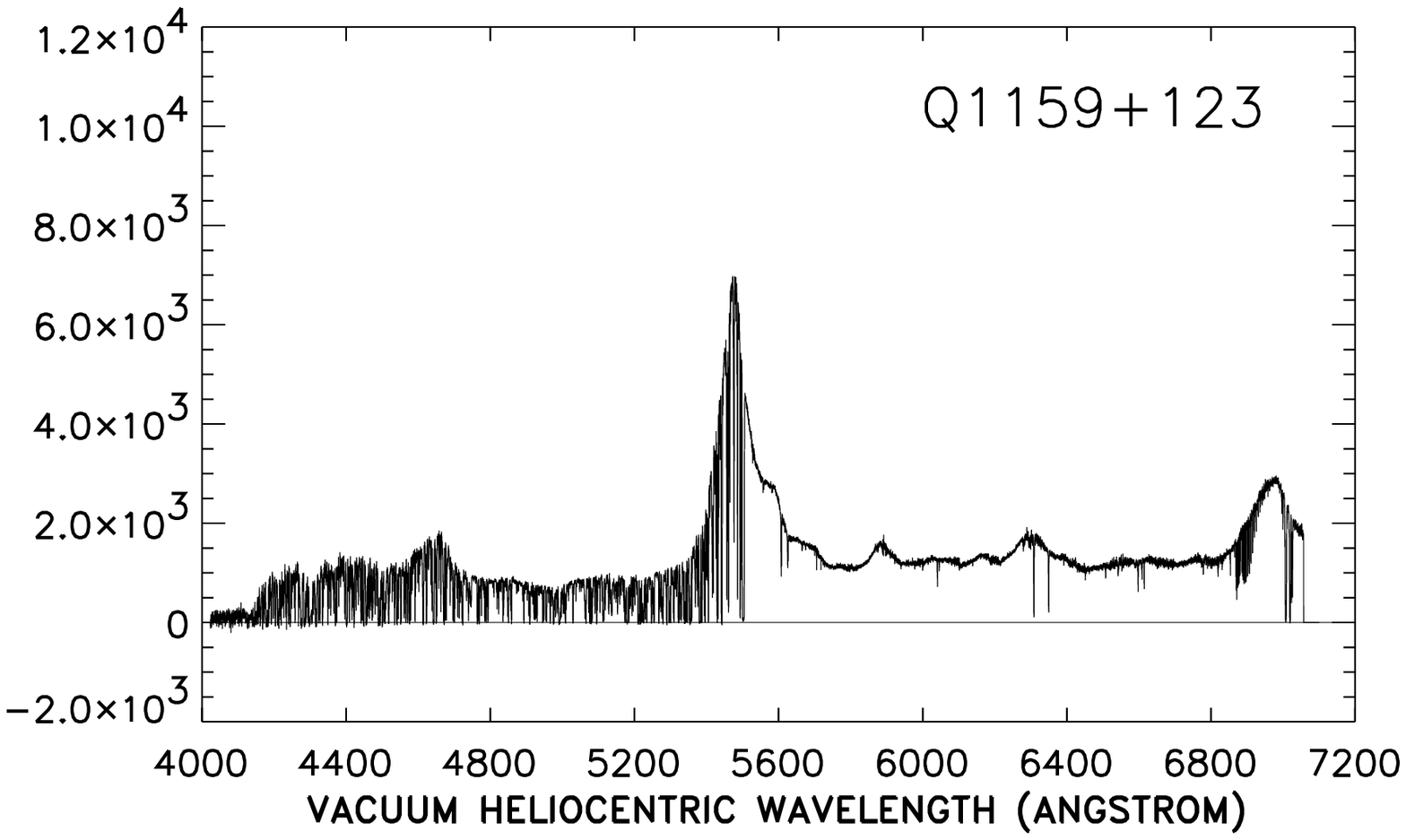}{5.5in}{0.0}{120}{120}{-300}{0}
\caption{Complete spectrum of Q1159+123 formed by
interpolating all the extracted orders to a uniform $0.05~{\rm \AA}$\
wavelength scale and then splicing adjacent orders.  The principal feature is
the quasar's ${\rm Ly}\alpha$\ emission near $5400~{\rm \AA}$\ and the
spectrum extends just past C~IV emission near $7000~{\rm \AA}$.  There is a
Lyman limit system near $4150~{\rm \AA}$.  Most of the features longward of
the ${\rm Ly}\alpha$\ emission are C~IV doublets, except for the atmospheric A
band.  
}\label{newfig:3}
\end{figure}

\begin{figure}
\figurenum{4}
\plotfiddle{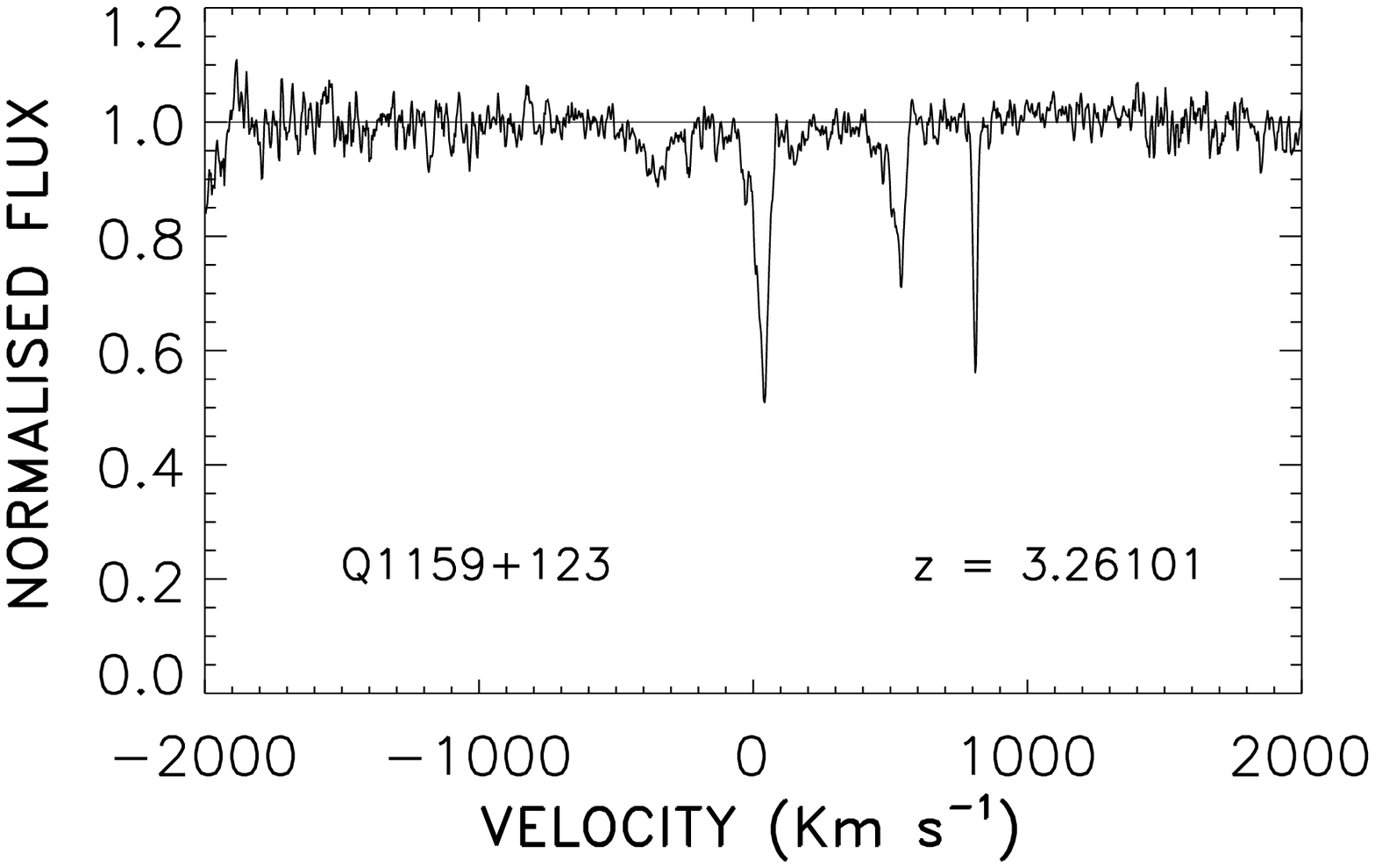}{5.5in}{0.0}{120}{120}{-300}{0}
\caption{A typical continuum-fitted region of spectrum,
around the $z = 3.26101$\ C~IV absorption line complex in Q1159+123.  The
continuum was fitted to a $\pm 2000~{\rm km\ s}^{-1}$\ region around the 
C~IV~$\lambda 1548~{\rm \AA}$\ line by iterating a 2nd-order polynomial fit to
a 10-pt smoothed spectrum and rejecting significantly deviant regions.
}\label{newfig:4}
\end{figure}

\begin{figure}
\figurenum{5a--i}
\plotfiddle{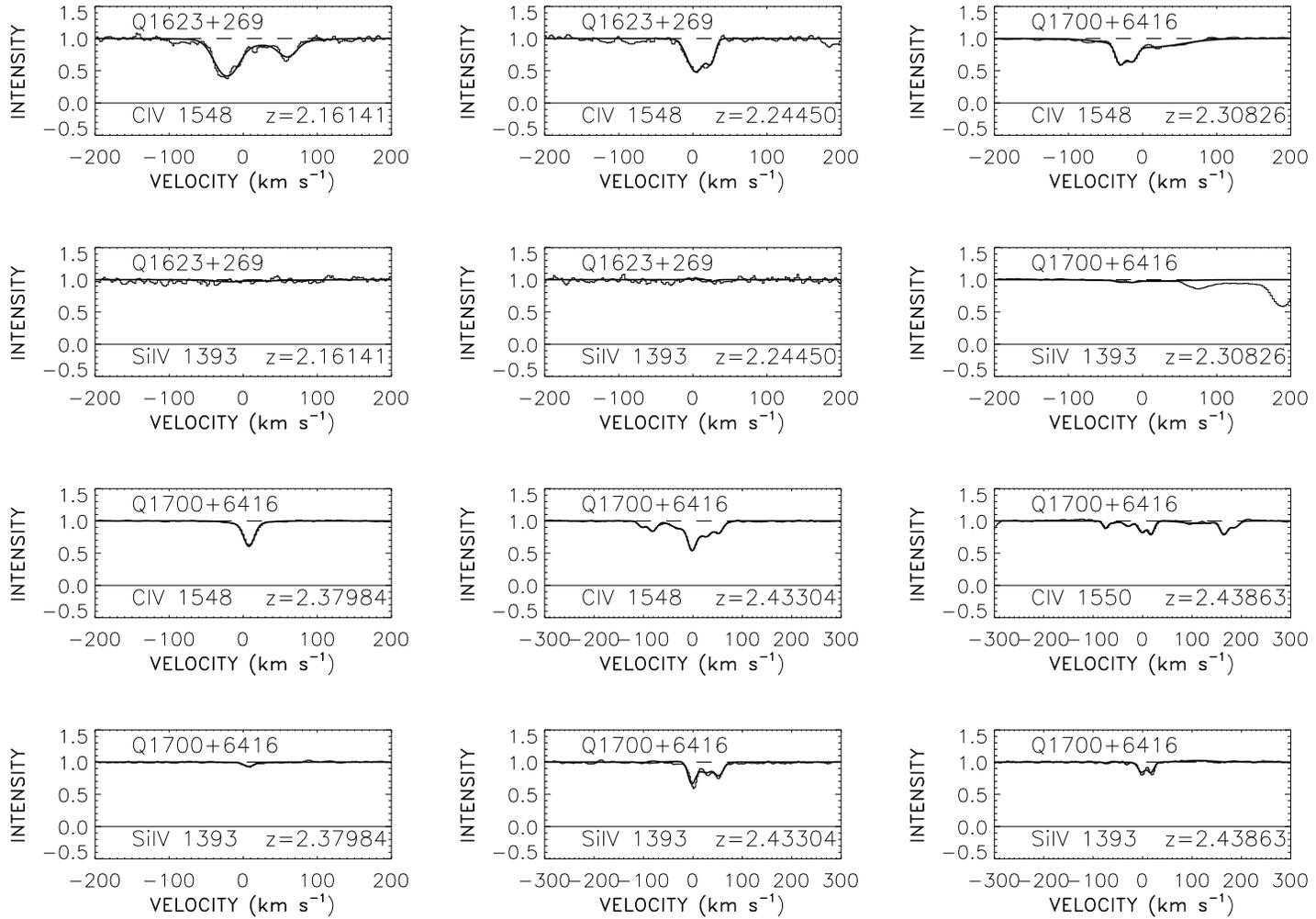}{5.5in}{0.0}{110}{110}{-262}{0}
\caption{Voigt profile fits to the C~IV and Si~IV 
doublets in all cloud complexes in the sample with $10^{13}~\rm cm^{-2} <
N({\rm C~IV}) < 10^{14}~\rm cm^{-2}$.  All doublets lie redward of the
quasar's ${\rm Ly}\alpha$\ emission.  Table~2 summarises the redshifts, number
of Voigt components, total C~IV and Si~IV column densities of each complex,
and C~II and Si~II column densities.  
}\label{fig:1}
\end{figure}

\begin{figure}
\figurenum{6}
\plotfiddle{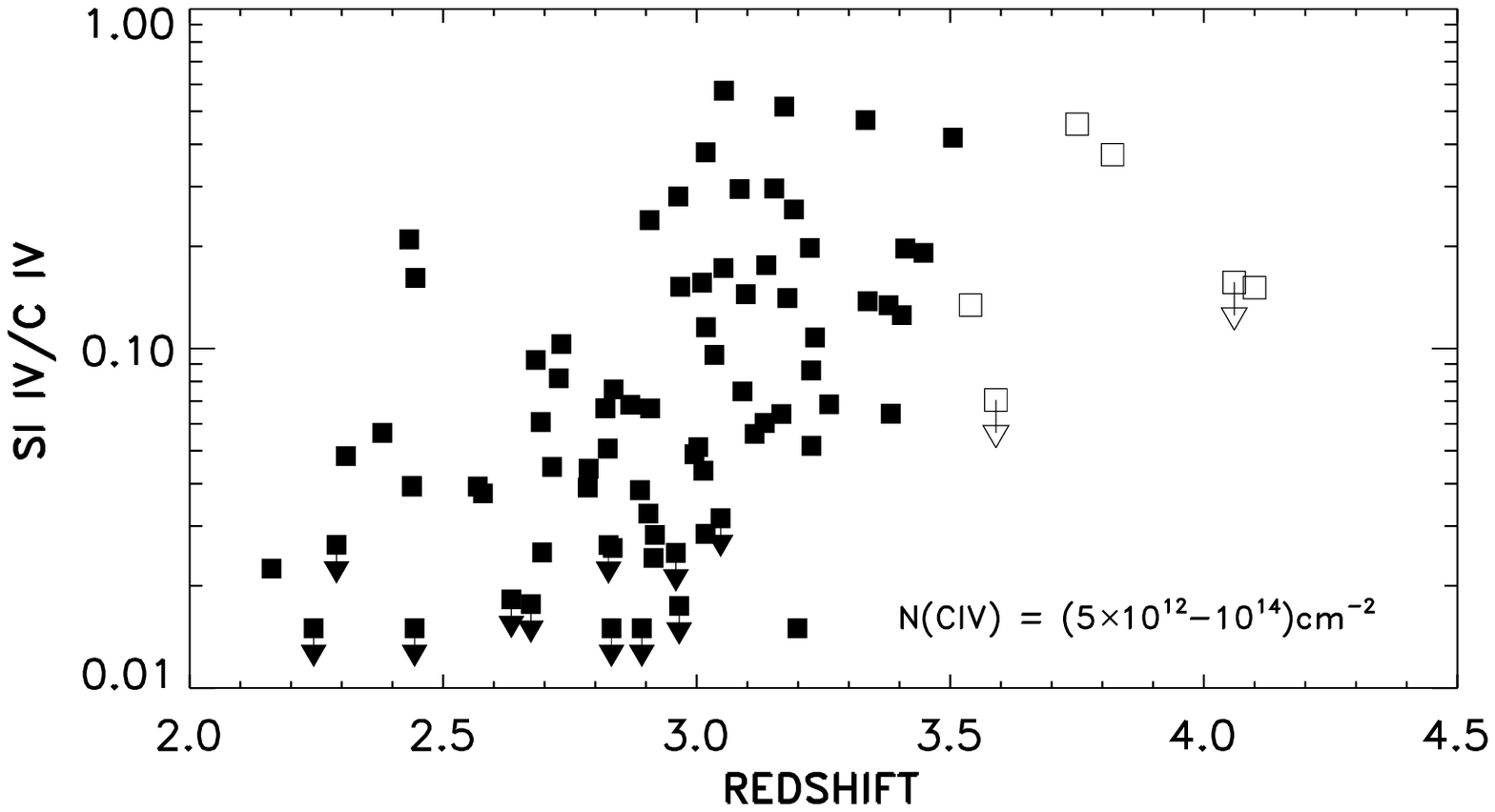}{5.5in}{0.0}{120}{120}{-300}{0}
\caption{The ratio of total Si~IV column density to total
C~IV column density as a function of redshift for the 76 complexes with
$5\times 10^{12}~\rm cm^{-2} \le N({\rm C~IV}) \le 10^{14}~\rm cm^{-2}$\
(filled squares).  Open squares denote values from Savaglio et al.'s (1997)
observations of Q0000$-$263.  Downward pointing arrows show systems in which
Si~IV is not detected and where the points are positioned at the $1~\sigma$\
upper limit.  
}\label{fig:2}
\end{figure}

\begin{figure}
\figurenum{7}
\plotfiddle{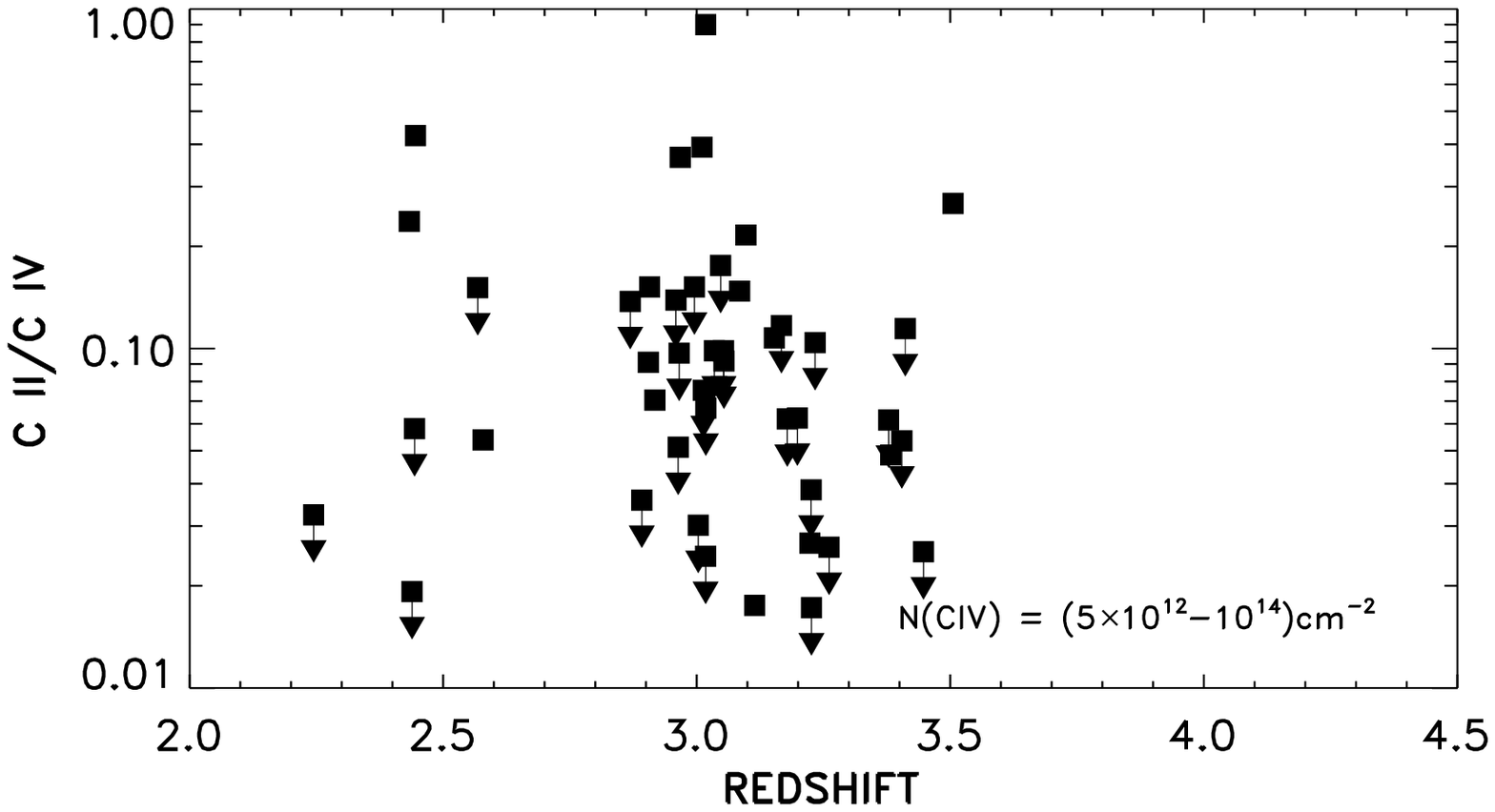}{5.5in}{0.0}{120}{120}{-300}{0}
\caption{The ratio of total C~II column density to total
C~IV column density as a function of redshift for the complexes of Figure~6 in
which the C~II system lies longer in wavelength than the forest.  Downward
pointing arrows show systems in which C~II is not detected and where the
points are positioned at the $1~\sigma$\ upper limit.  
}\label{fig:3}
\end{figure}

\begin{figure}
\figurenum{8}
\plotfiddle{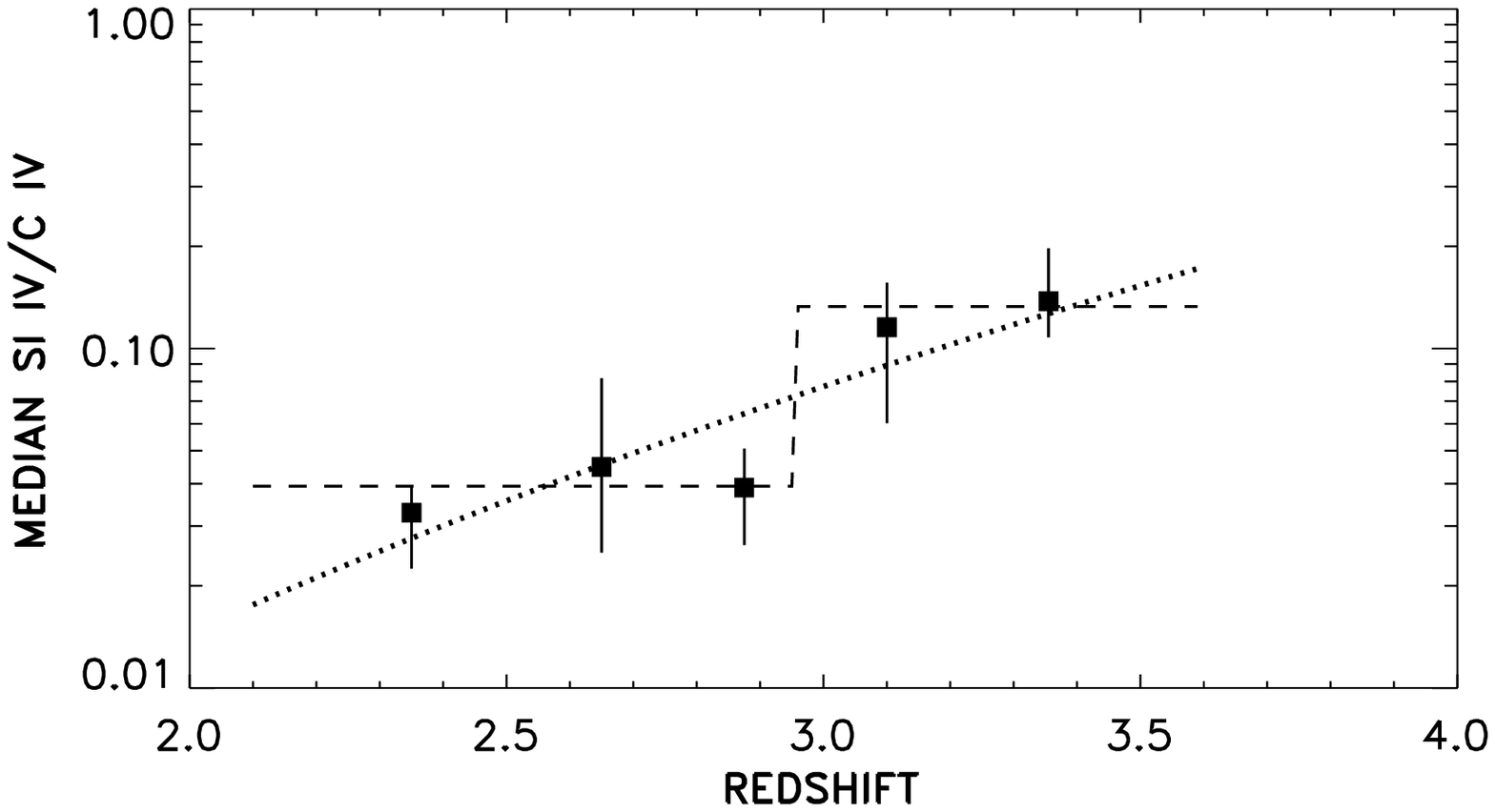}{5.5in}{0.0}{120}{120}{-300}{0}
\caption{Median values of Si~IV/C~IV as a function of
redshift for the sample of Figure~6.  Errors are $\pm 1~\sigma$\ computed
using the median sign method.  The median value of Si~IV/C~IV is
$0.039^{+0.009}_{-0.006}$\ for $z < 3$\ and $0.13^{+0.04}_{-0.04}$\ for $z >
3$.  The dashed line is a step function fit to the data, with a jump of a
factor of 3.4 at $z = 2.95$.  The dotted line has the form $(1+z)^{5.4}$.  It
does not predict a sufficient jump between $z = 2.85$\ and $z = 3.1$\ to
reconcile the low and high redshift distributions (Figure~9).
}\label{fig:4}
\end{figure}

\begin{figure}
\figurenum{9}
\plotfiddle{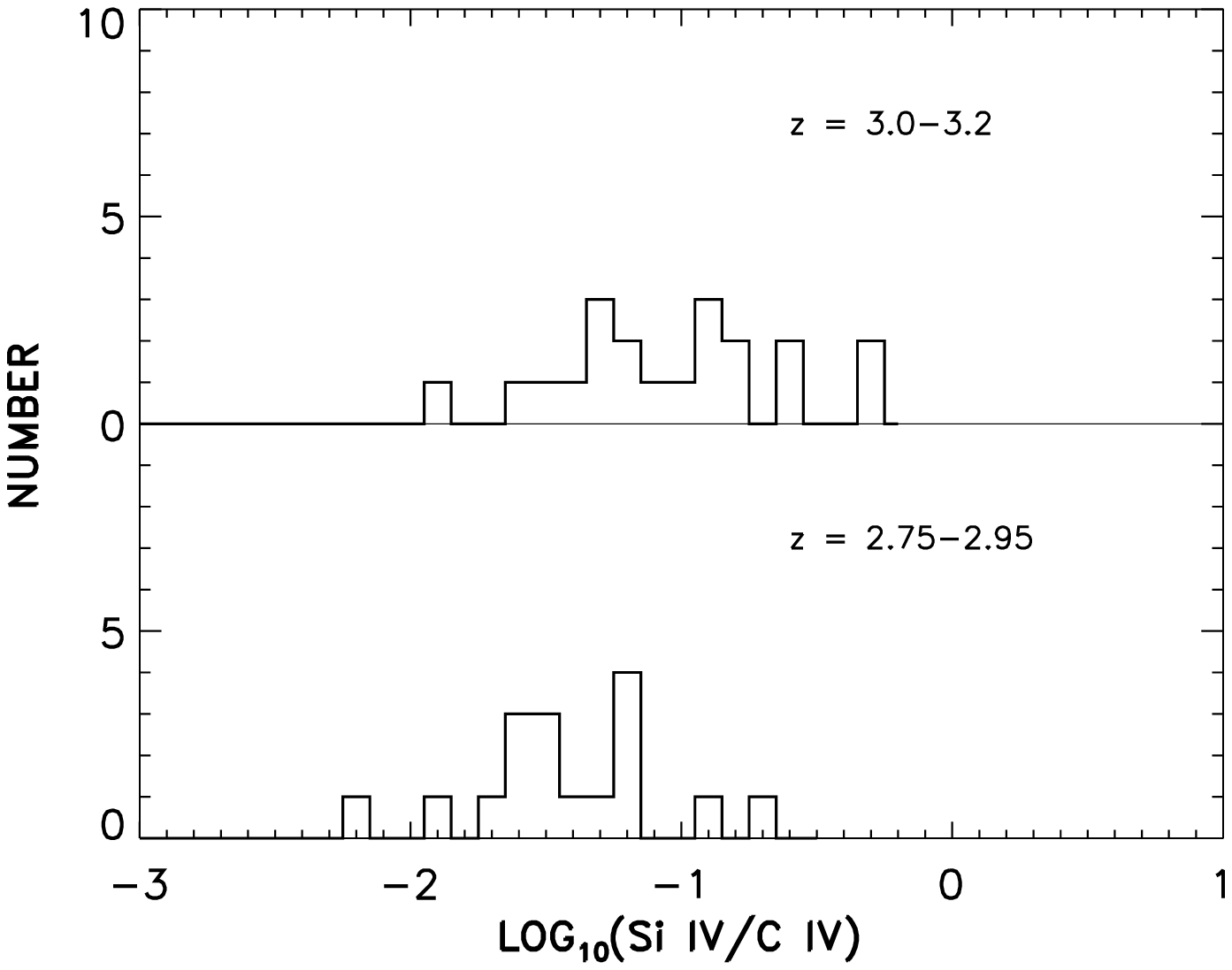}{5.5in}{0.0}{120}{120}{-300}{0}
\caption{Histograms of the distributions of $\log_{10}({\rm
Si~IV/C~IV})$\ for 22 systems with $z = 3.0 - 3.2$\ (top) with a median of
$0.14^{+0.04}_{-0.075}$\ and 16 systems with $z = 2.75 - 2.95$\ (bottom) with
a median of $0.038^{+0.029}_{-0.010}$.  A multiplicative jump of a least a
factor of 1.8 is required at the 98\% confidence level to reconcile the low
and high redshift distributions.
}\label{fig:5}
\end{figure}

\clearpage
\newpage

\begin{figure}
\figurenum{10}
\plotfiddle{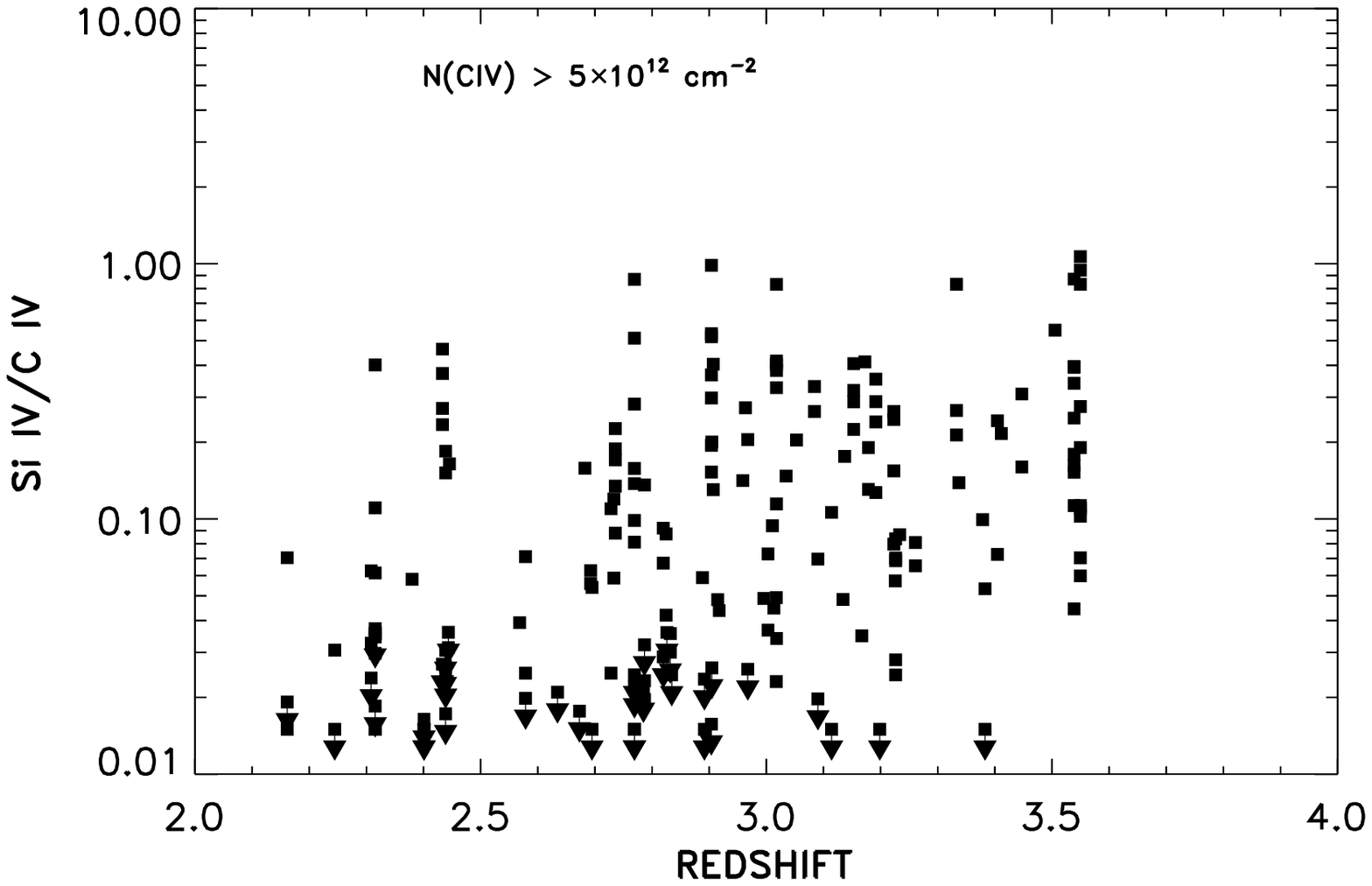}{5.5in}{0.0}{120}{120}{-300}{0}
\caption{Si~IV/C~IV as a function of redshift for all Voigt
profile components with $N({\rm C~IV}) > 5\times 10^{12}~\rm cm^{-2}$\ in the
sample, irrespective of the total column density of the complex, and including
a number of Lyman limit systems.  
}\label{fig:6}
\end{figure}

\begin{figure}
\figurenum{11}
\plotfiddle{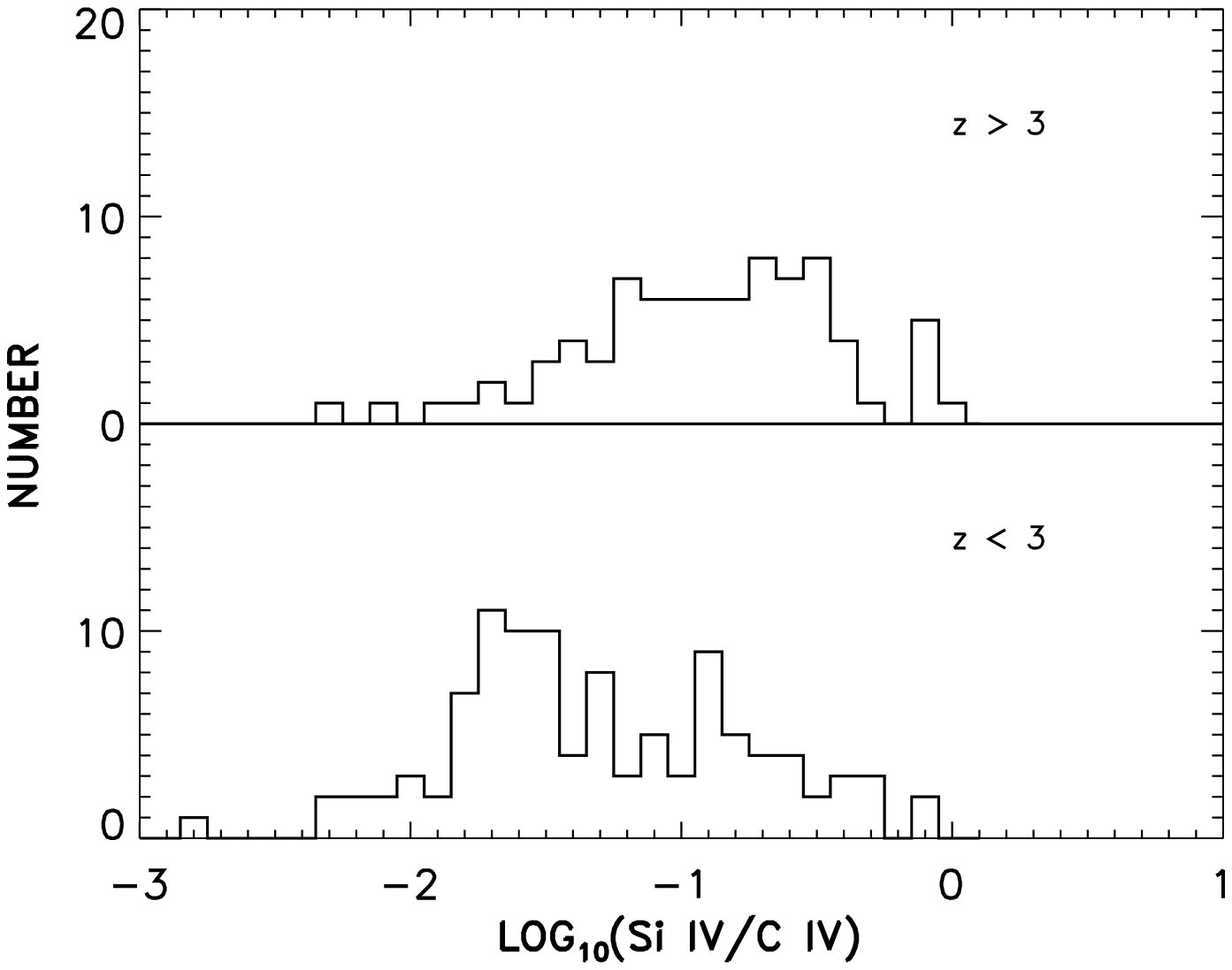}{5.5in}{0.0}{120}{120}{-300}{0}
\caption{Histograms of the distributions of $\log_{10}({\rm
Si~IV/C~IV})$\ for all Voigt profile components in the sample of Figure~10
with $z > 3$\ (top) with a median of $0.15^{+0.04}_{-0.04}$\ and for $z < 3$\
(bottom) with a median of $0.043^{+0.015}_{-0.008}$.  A multiplicative jump of
a least a factor of 1.93 is required at the 98\% confidence level to reconcile
the low and high redshift distributions.  
}\label{fig:7}
\end{figure}

\begin{figure}
\figurenum{12}
\plotfiddle{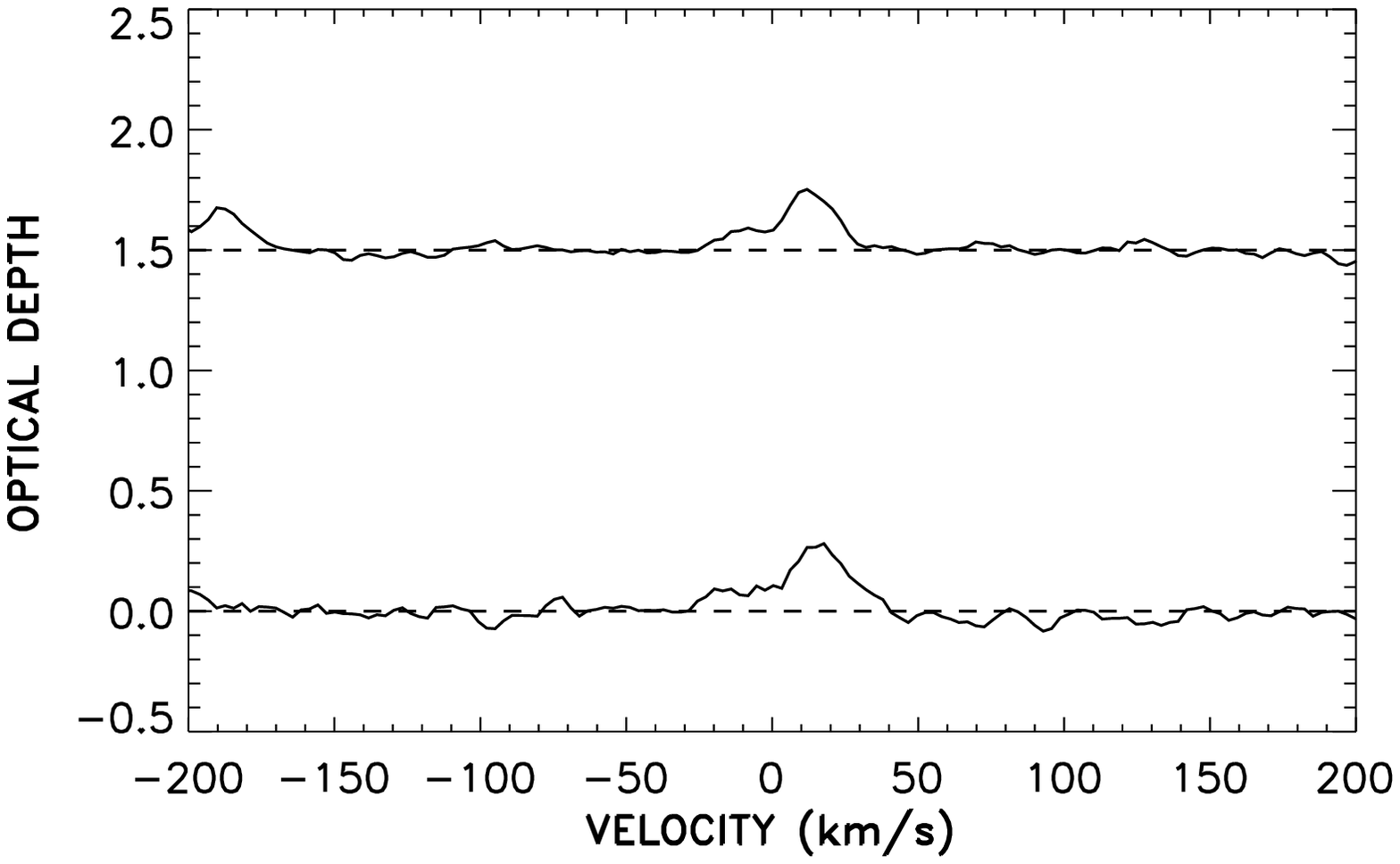}{5.5in}{0.0}{120}{120}{-300}{0}
\caption{Sample optical depths of Si~IV (top) and C~IV
(bottom) for the $z = 3.33714$\ complex in Q2000$-$330.  The Si~IV optical
depth has been offset by 1.5.
}\label{fig:8}
\end{figure}

\begin{figure}
\figurenum{13}
\plotfiddle{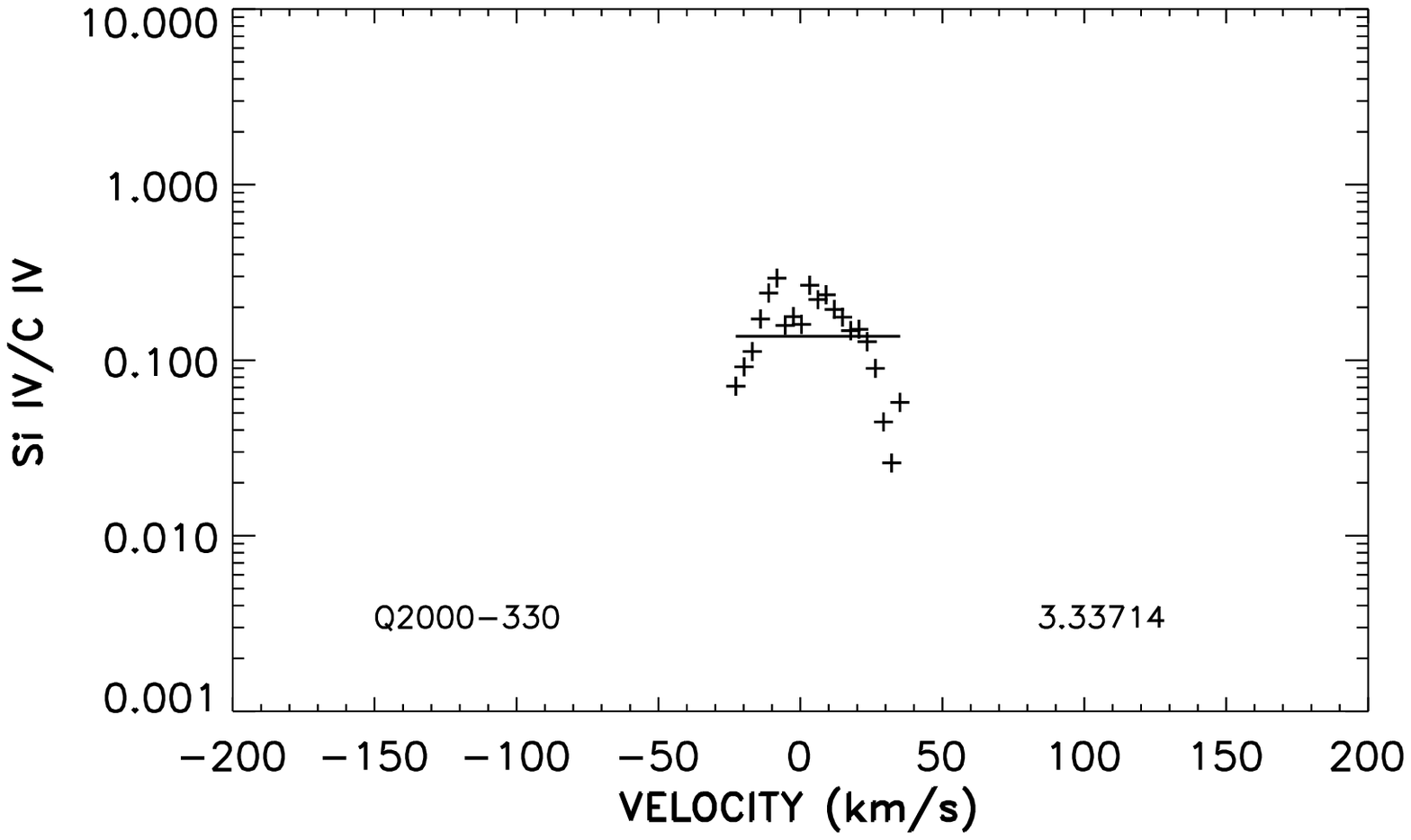}{5.5in}{0.0}{120}{120}{-300}{0}
\caption{Si~IV/C~IV values across the $z = 3.33714$\ complex
in Q2000$-$330, computed from the Si~IV and C~IV optical depths of Figure~12
with $\tau({\rm C~IV}) > 0.05$, as $N({\rm ion}) = \tau\lambda f$, where
$\lambda$\ is the wavelength and $f$\ the oscillator strength of the
transition.  The solid line is the average value of Si~IV/C~IV in the complex,
computed from the total column densities of Si~IV and C~IV measured from Voigt
profile fitting.
}\label{fig:9}
\end{figure}

\begin{figure}
\figurenum{14}
\plotfiddle{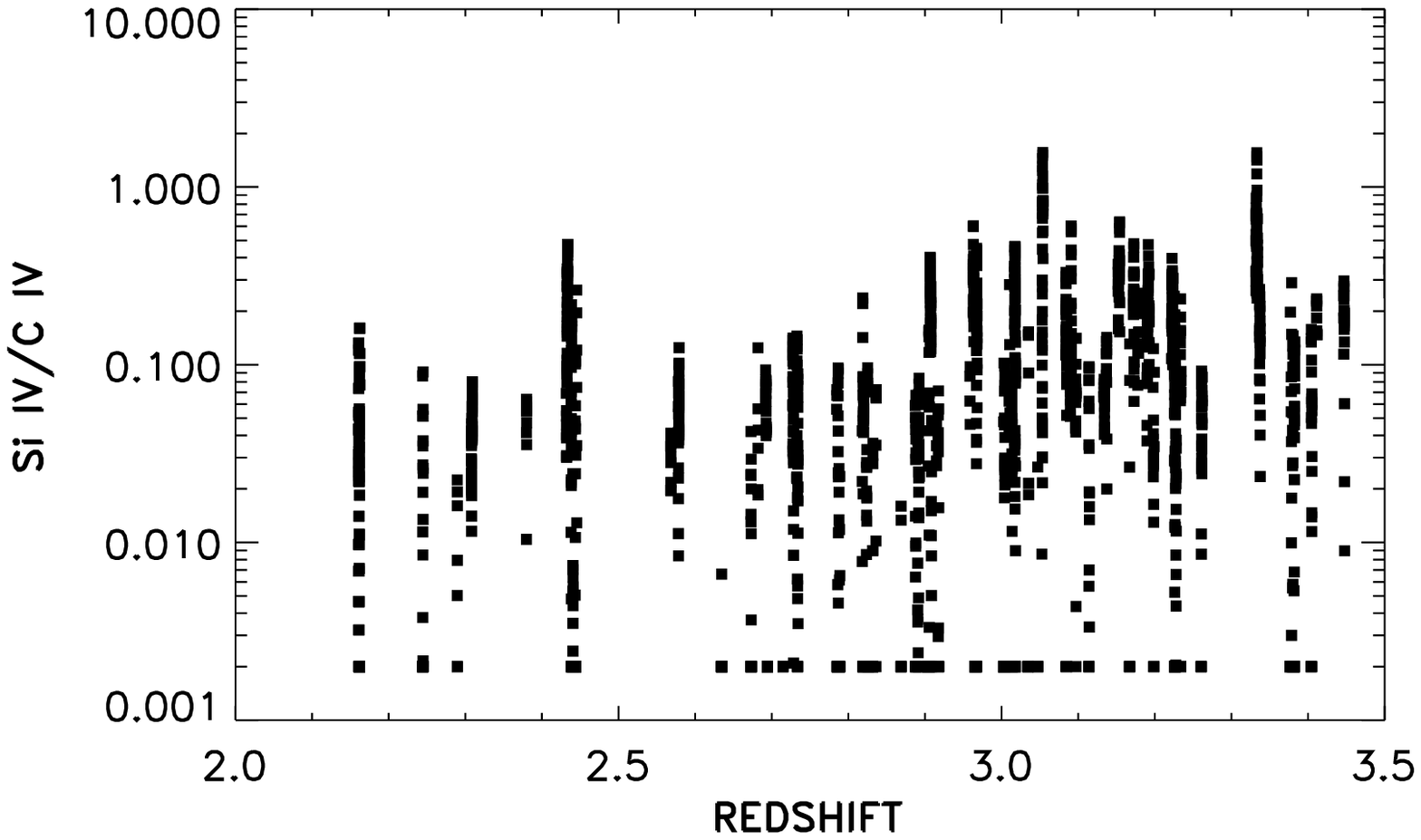}{5.5in}{0.0}{120}{120}{-300}{0}
\caption{The Si~IV/C~IV density ratios computed from
optical depths as in Figure~13 for all the absorption line complexes of
Table~2, as a function of the redshift of the complex.  
}\label{fig:10}
\end{figure}

\begin{figure}
\figurenum{15a--c}
\plotfiddle{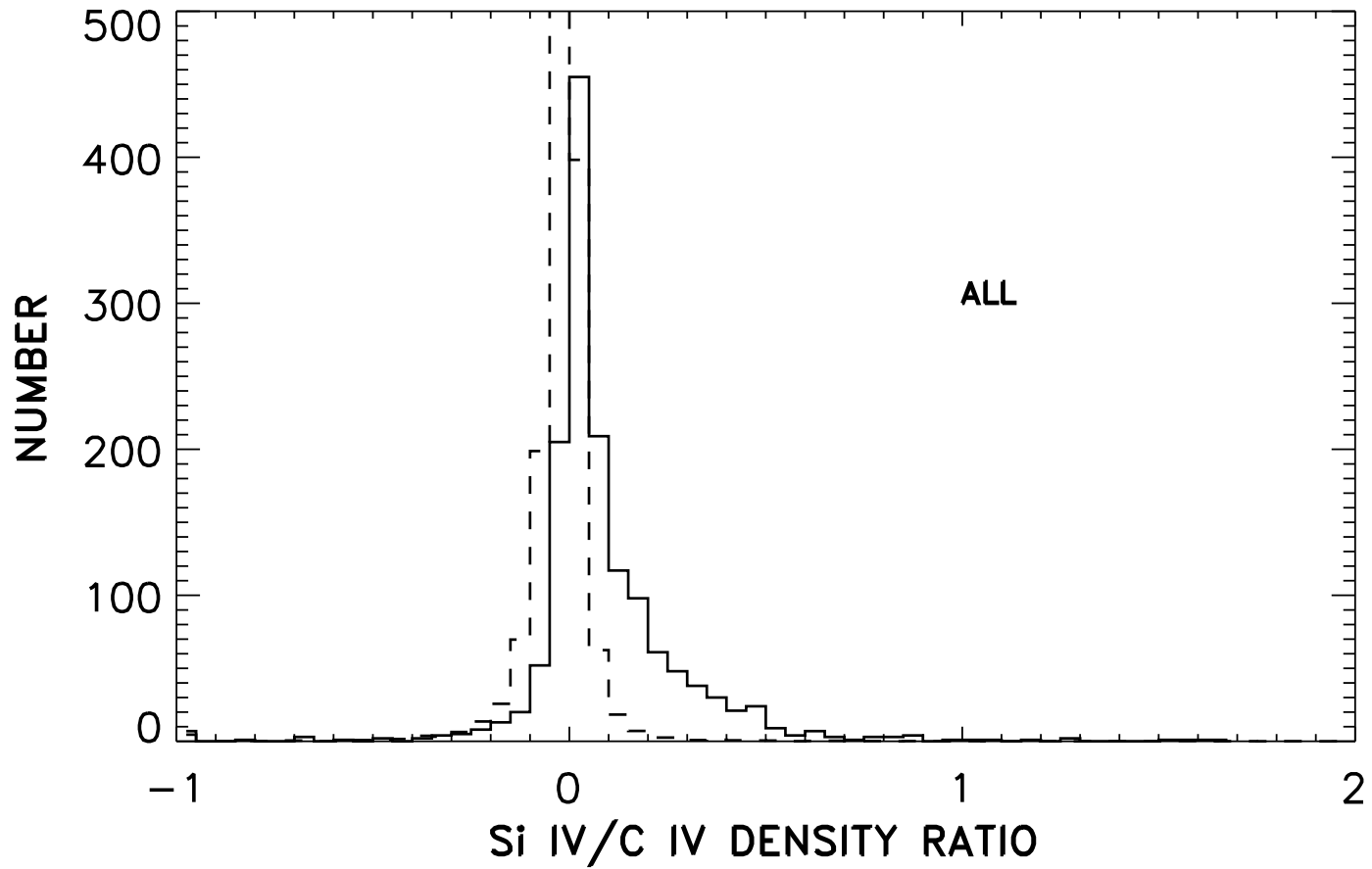}{5.5in}{0.0}{120}{120}{-300}{0}
\caption{The distribution of Si~IV/C~IV density ratios
for all the data of Figure~14 (solid line) compared to the average of random
realizations (dashed line).  Median and mean values and $1~\sigma$\ errors are
given in Tables~3 and 4, respectively. 
}\label{fig:11}
\end{figure}

\begin{figure}
\figurenum{15b}
\plotfiddle{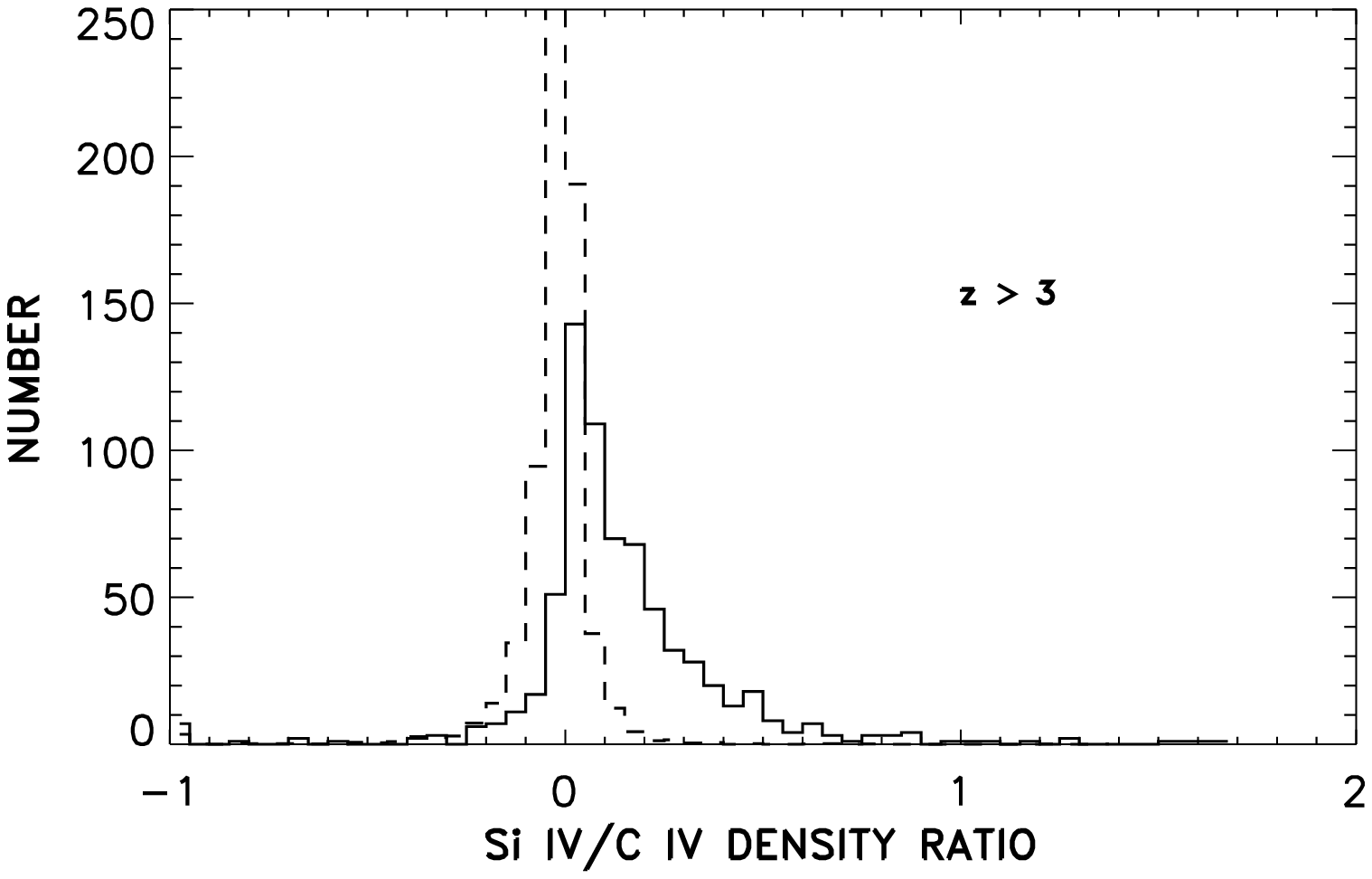}{5.5in}{0.0}{120}{120}{-300}{0}
\caption{As in (a) for $z > 3$. 
}
\end{figure}

\begin{figure}
\figurenum{15c}
\plotfiddle{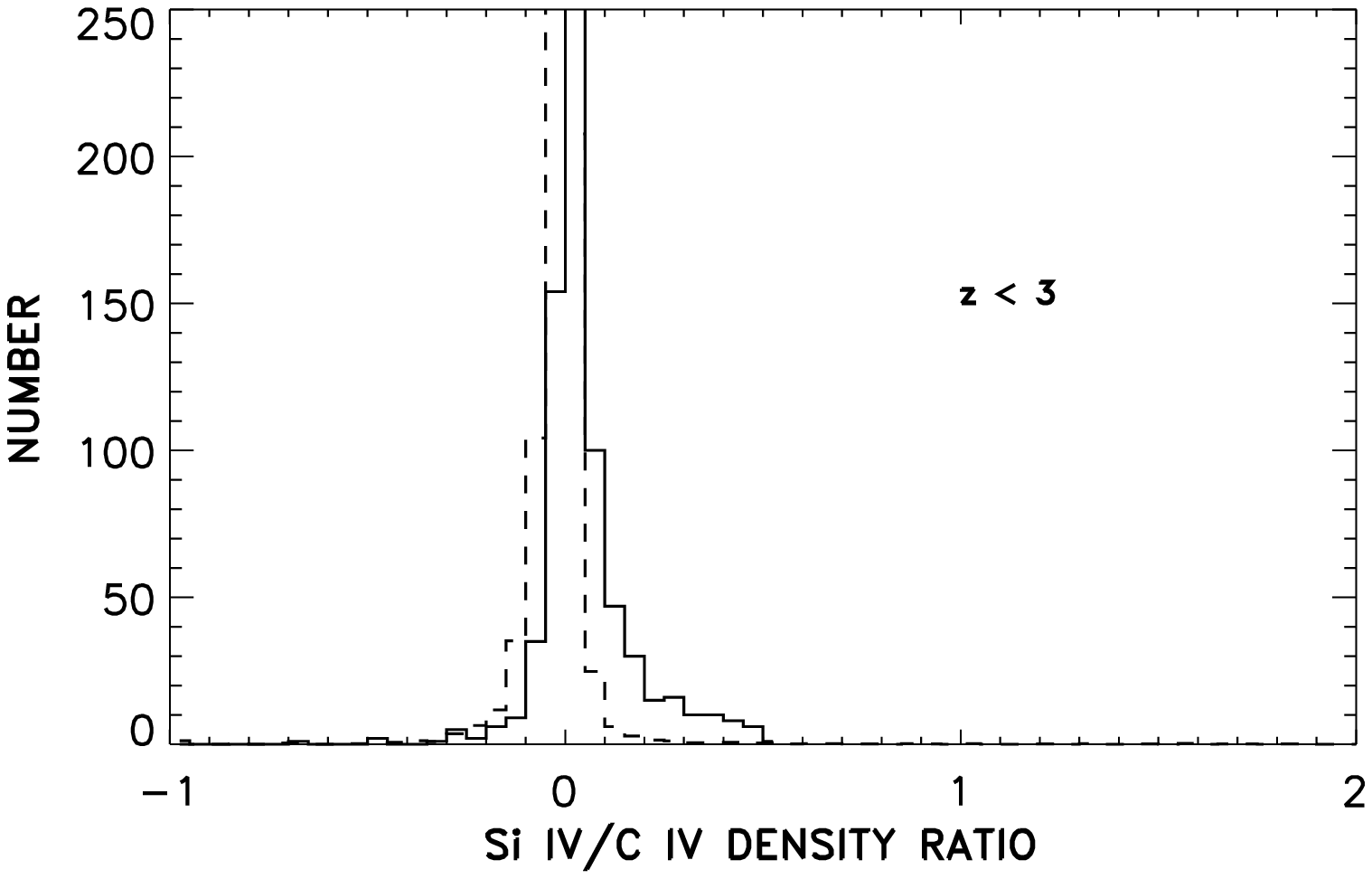}{5.5in}{0.0}{120}{120}{-300}{0}
\caption{
As in (a) for $z < 3$.  
}
\end{figure}

\begin{figure}
\figurenum{16}
\plotfiddle{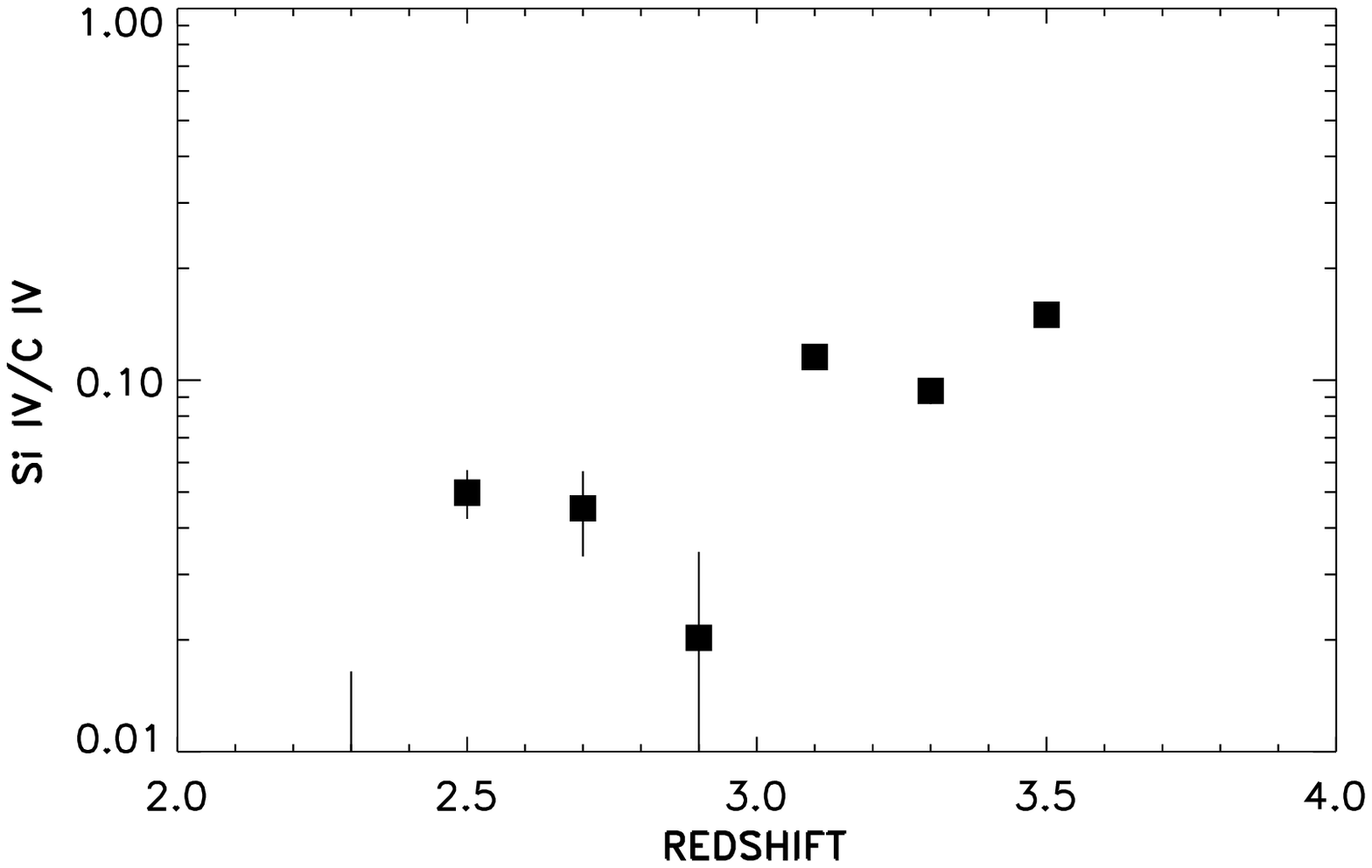}{5.5in}{0.0}{120}{120}{-300}{0}
\caption{Median Si~IV/C~IV as a function of redshift for
the data of Figure~14, with $\pm 1~\sigma$\ errors.  The median is $0.035 \pm
0.005$\ for $z < 3$\ and $0.10 \pm 0.01$\ for $z > 3$.
}\label{fig:12}
\end{figure}

\begin{figure}
\figurenum{17}
\plotfiddle{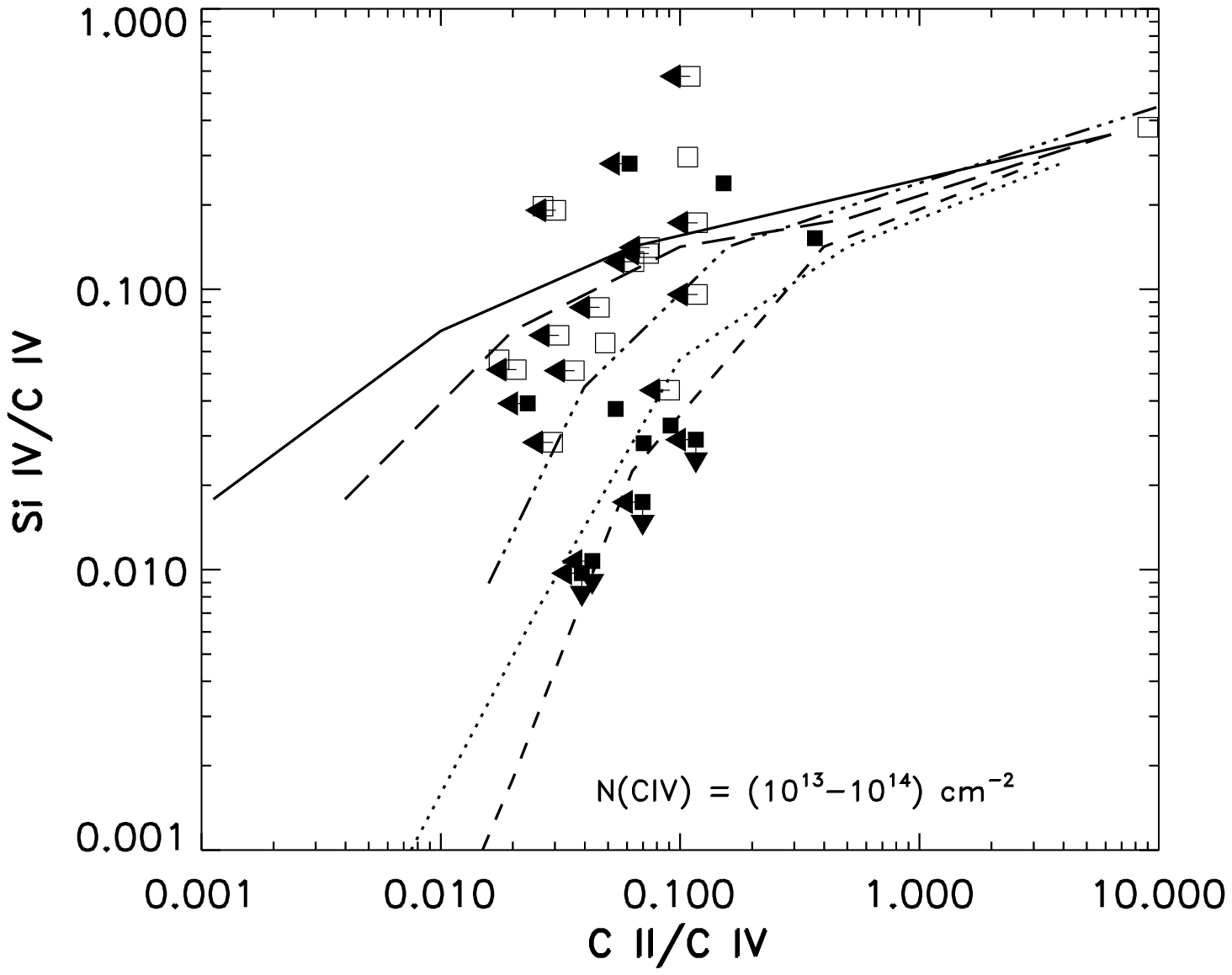}{5.5in}{0.0}{120}{120}{-300}{0}
\caption{Si~IV/C~IV versus C~II/C~IV for complexes
with $10^{13}~\rm cm^{-2} < N({\rm C~IV}) < 10^{14}~\rm cm^{-2}$\ and with C~II longward of
the forest for systems with $z < 3$\ (filled squares) and $z > 3$\ (open
squares).  Arrows mark systems in which Si~IV or C~II is not detected, with
the square positioned at the $1~\sigma$\ level.  The curves are models
computed with the CLOUDY code (Ferland 1993) with
$\log_{10}({\rm Si/C}) = -0.66$\ for a $-1.5$\ power law ionizing spectrum
(dashed line) and with the addition of a break at the He$^+$\ edge at 4~Ryd of
a factor of 2 (dotted line), 10 (dash--dot line), 100 (long dash line) and
1000 (solid line). 
}\label{fig:13}
\end{figure}

\begin{figure}
\figurenum{18}
\plotfiddle{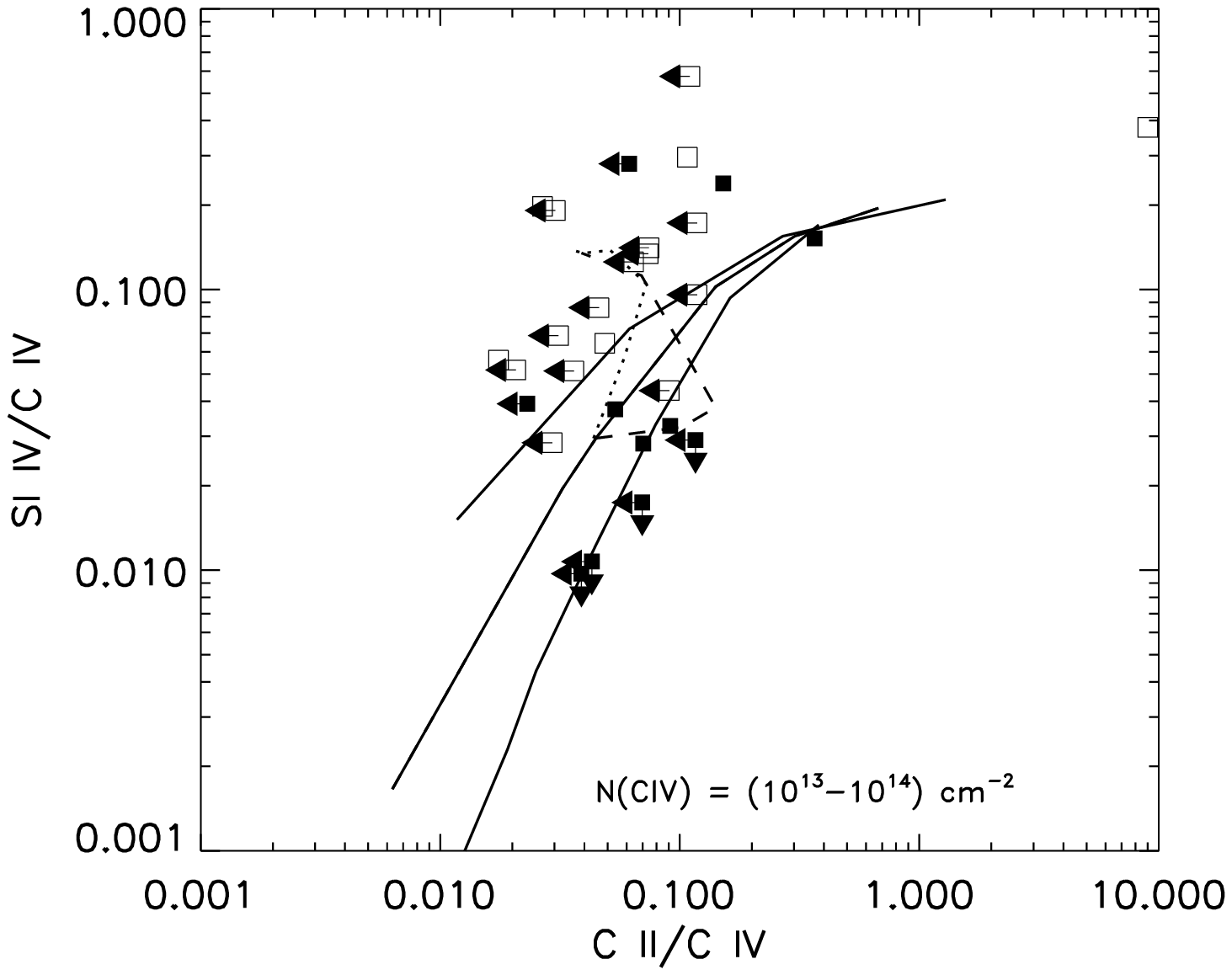}{5.5in}{0.0}{120}{120}{-300}{0}
\caption{Si~IV/C~IV versus C~II/C~IV for the data of
Figure~17 (filled and open squares).  The solid lines are CLOUDY models with
power law spectra with index $-1.5$\ (bottom), $-2.0$\ (middle), and $-2.5$\
(top), computed with $\log_{10}({\rm Si/C}) = -0.66$.  The dashed and dotted
curves show the effect of the introduction of a break in the spectrum at the
He$^+$\ edge.  See text (\S 4.2) for details.
}\label{fig:14}
\end{figure}

\clearpage
\newpage

\begin{figure}
\figurenum{19a}
\plotfiddle{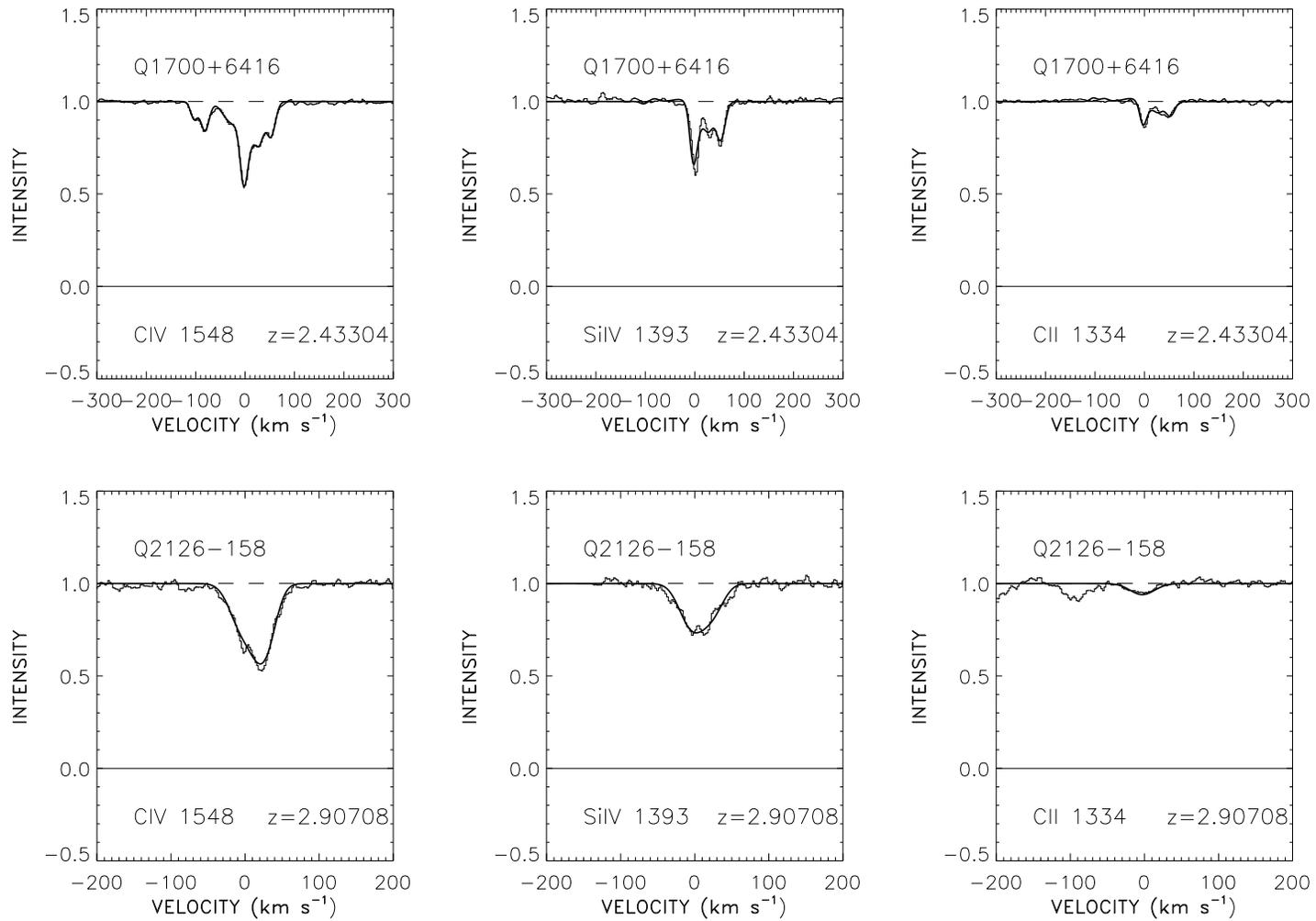}{5.5in}{0.0}{110}{110}{-262}{0}
\caption{Sample C~IV, Si~IV and C~II profiles for
systems with a high value of Si~IV/C~IV at $z < 3$.
}\label{fig:15}
\end{figure}

\begin{figure}
\figurenum{19b}
\plotfiddle{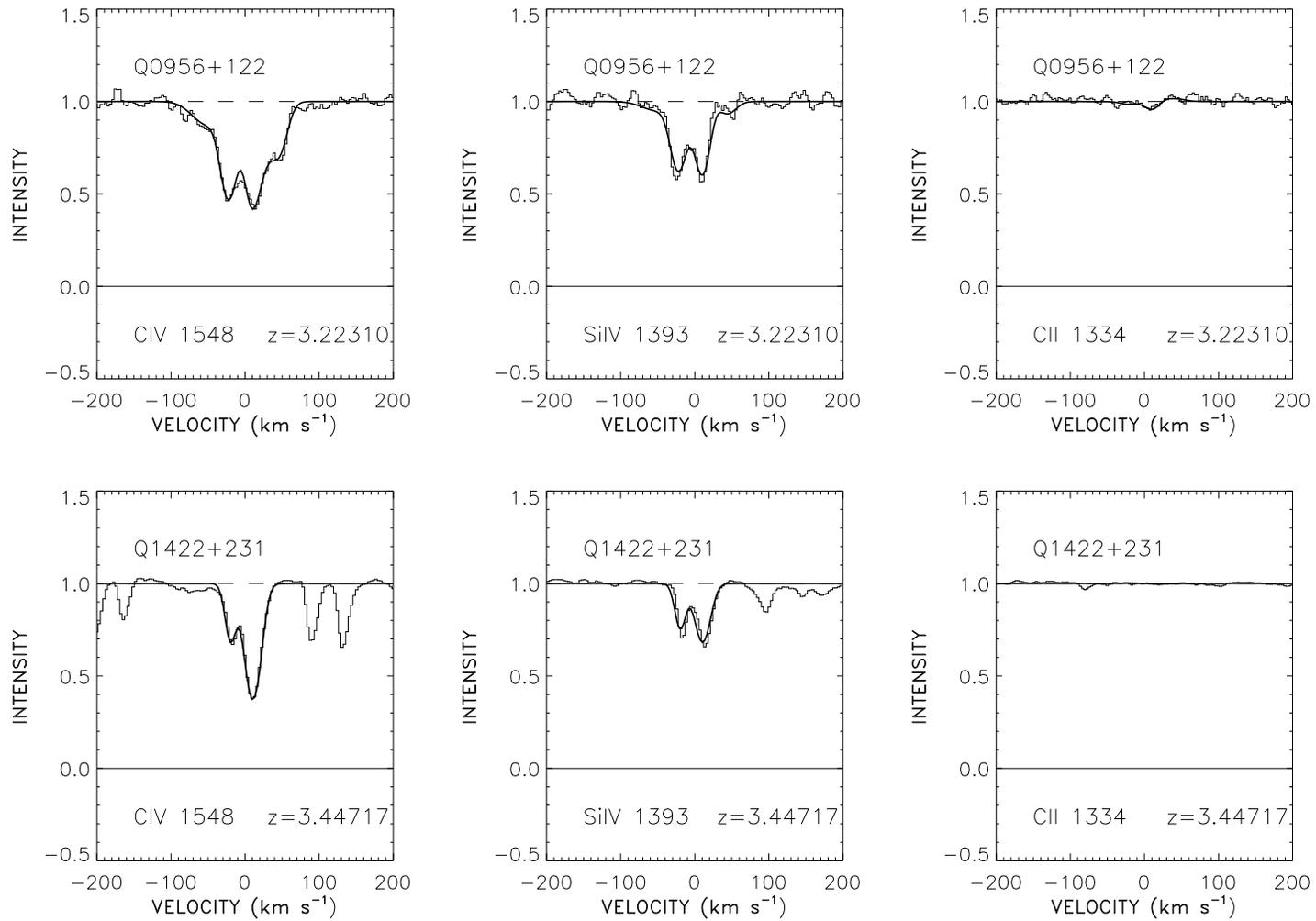}{5.5in}{0.0}{110}{110}{-262}{0}
\caption{As in (a) for
systems with $z > 3$, illustrating the absence of C~II in these high redshift
systems. 
}
\end{figure}

\begin{figure}
\figurenum{20a--g}
\plotfiddle{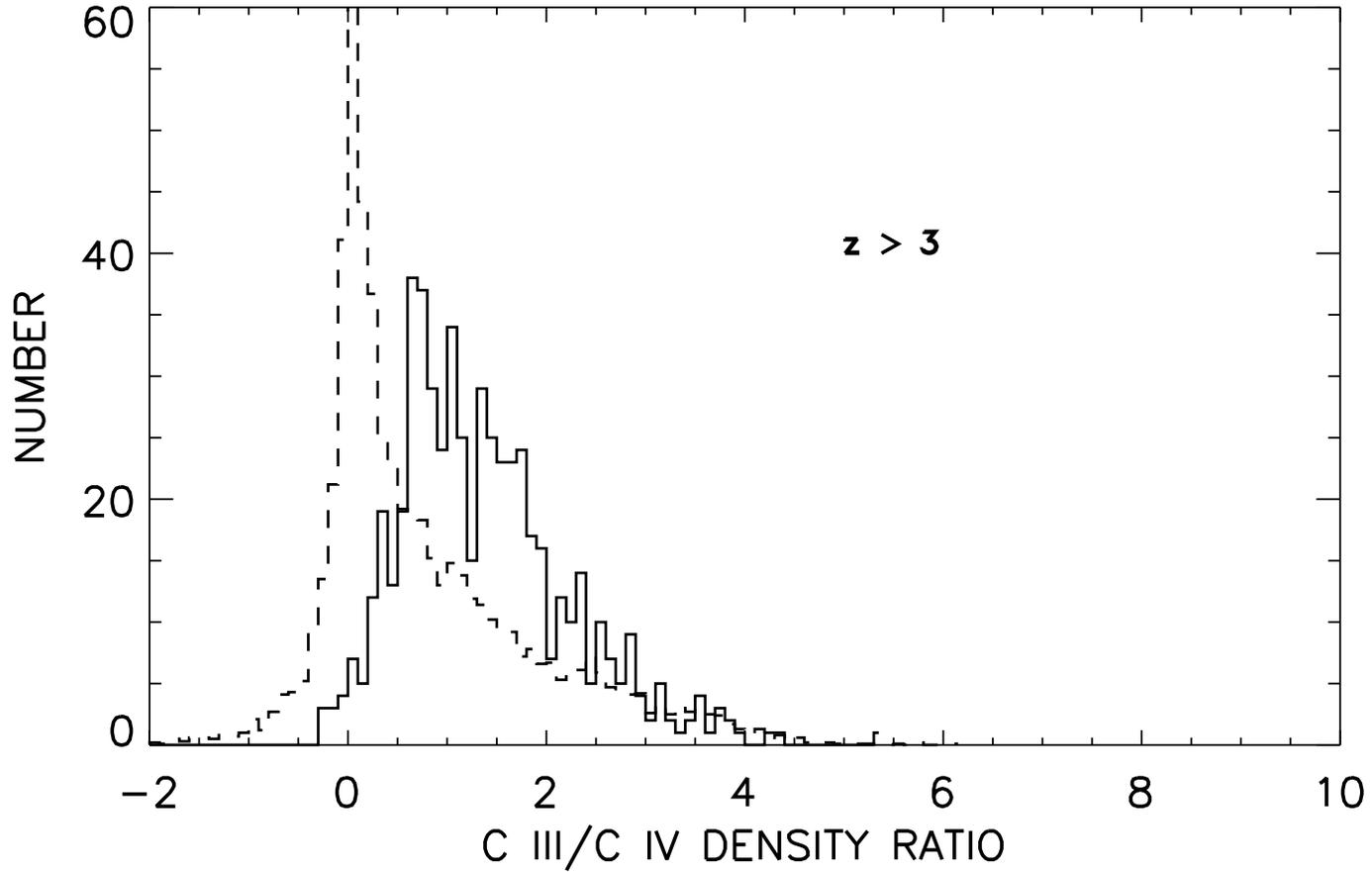}{5.5in}{0.0}{120}{120}{-300}{0}
\caption{(Solid lines): distributions of
C~III/C~IV, N~V/C~IV, and O~VI/C~IV computed from the optical depth ratios for
$\tau({\rm C~IV}) > 0.05$, as in Figure~15, compared with average blank
realizations (dashed lines).  Where possible, distributions are also shown
separately for $z > 3$\ and $z < 3$.  Median and mean values and $1~\sigma$\
errors are given in Tables~3 and 4, respectively.
}\label{fig:16}
\end{figure}

\begin{figure}
\figurenum{20b}
\plotfiddle{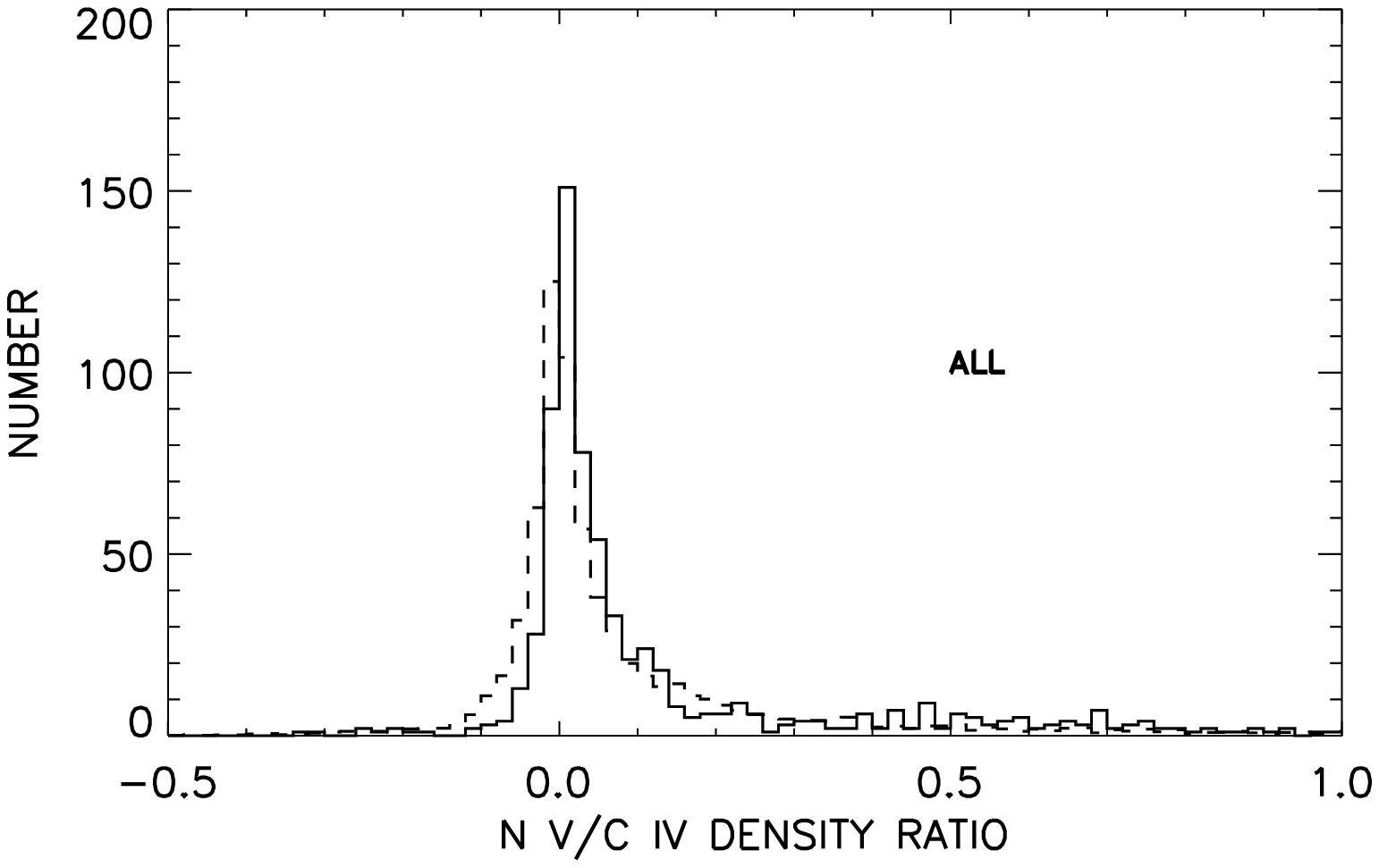}{5.5in}{0.0}{120}{120}{-300}{0}
\caption{
}
\end{figure}

\begin{figure}
\figurenum{20c}
\plotfiddle{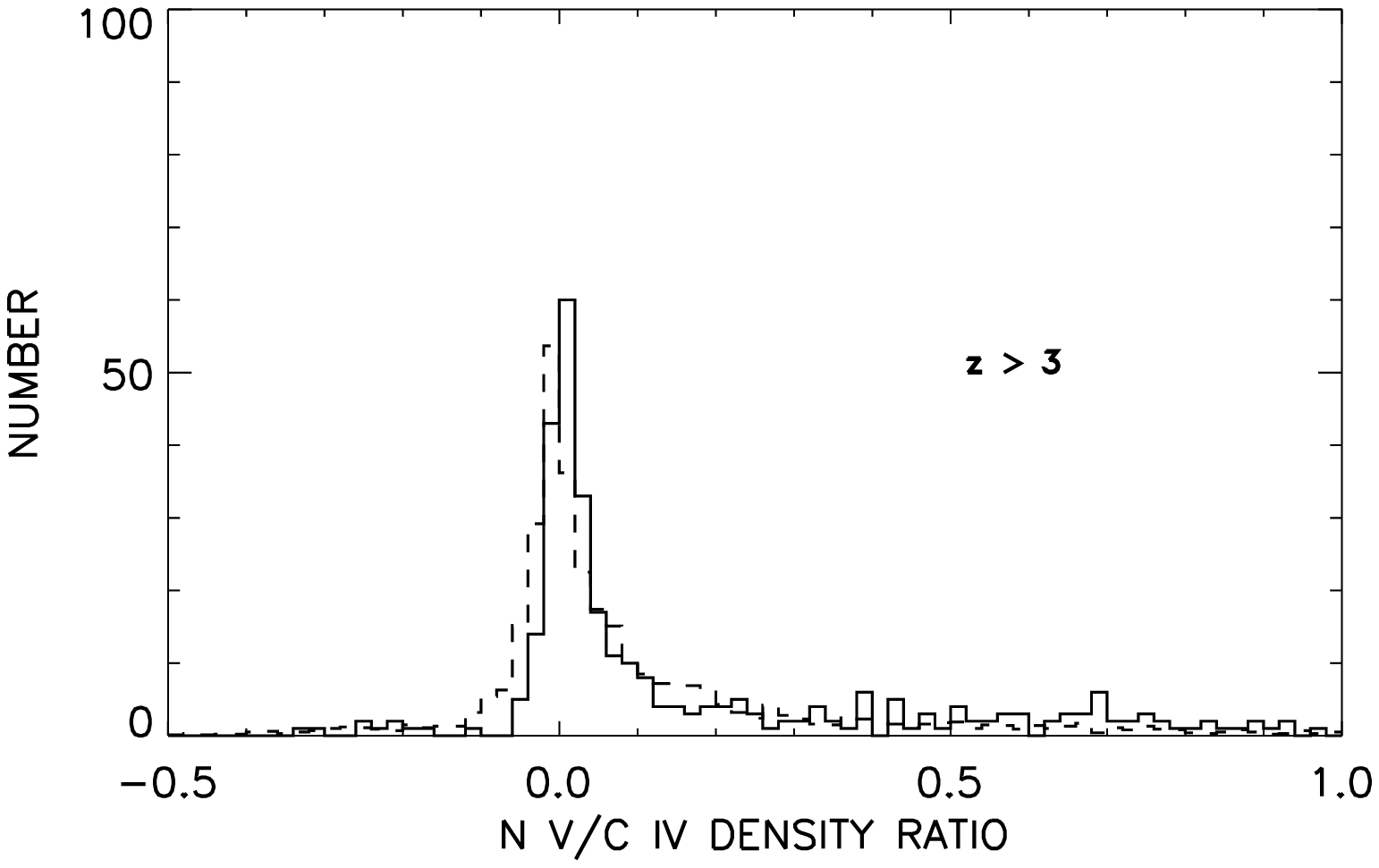}{5.5in}{0.0}{120}{120}{-300}{0}
\caption{
}
\end{figure}

\begin{figure}
\figurenum{20d}
\plotfiddle{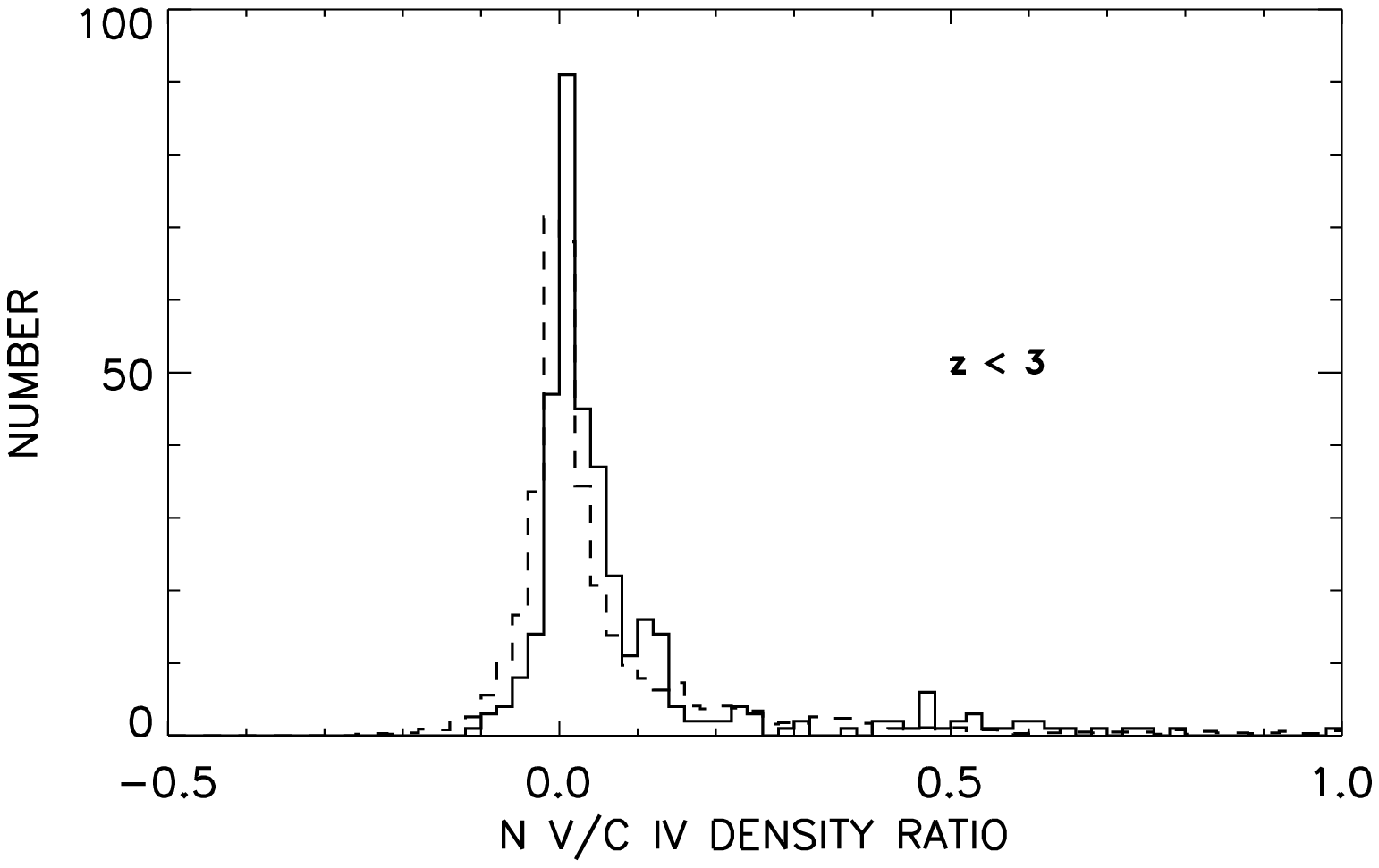}{5.5in}{0.0}{120}{120}{-300}{0}
\caption{
}
\end{figure}

\begin{figure}
\figurenum{20e}
\plotfiddle{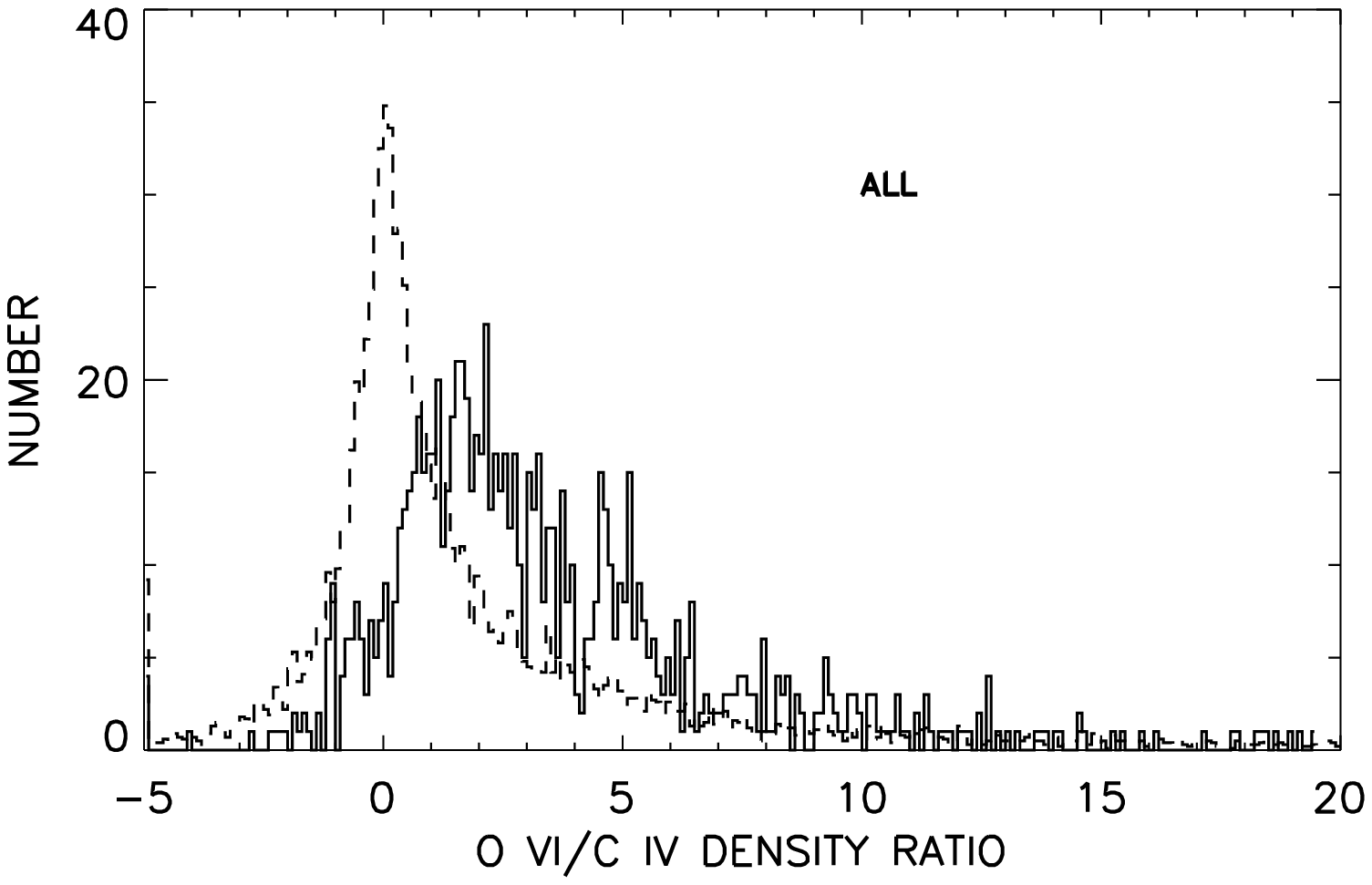}{5.5in}{0.0}{120}{120}{-300}{0}
\caption{
}
\end{figure}

\begin{figure}
\figurenum{20f}
\plotfiddle{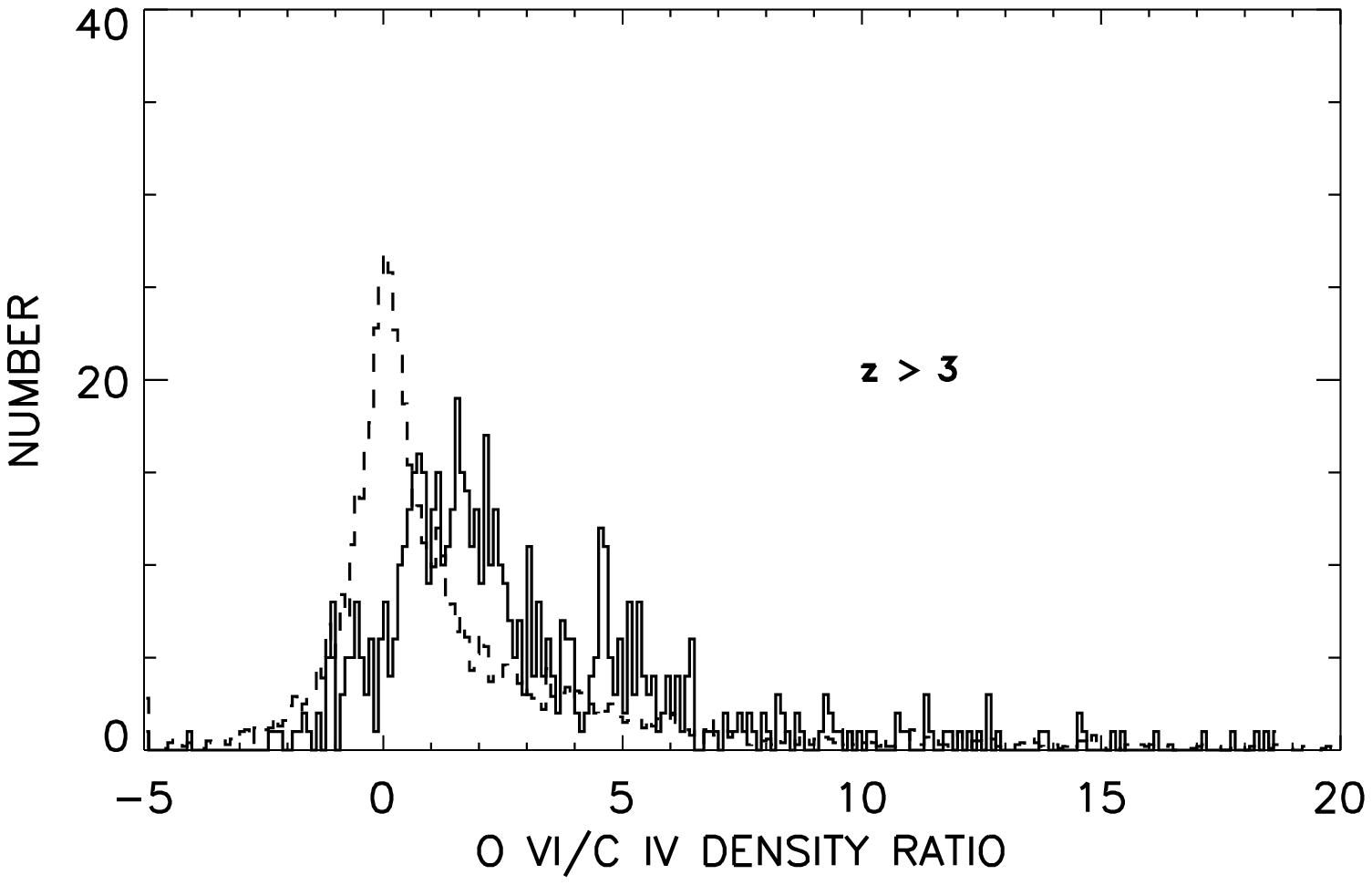}{5.5in}{0.0}{120}{120}{-300}{0}
\caption{
}
\end{figure}

\begin{figure}
\figurenum{20g}
\plotfiddle{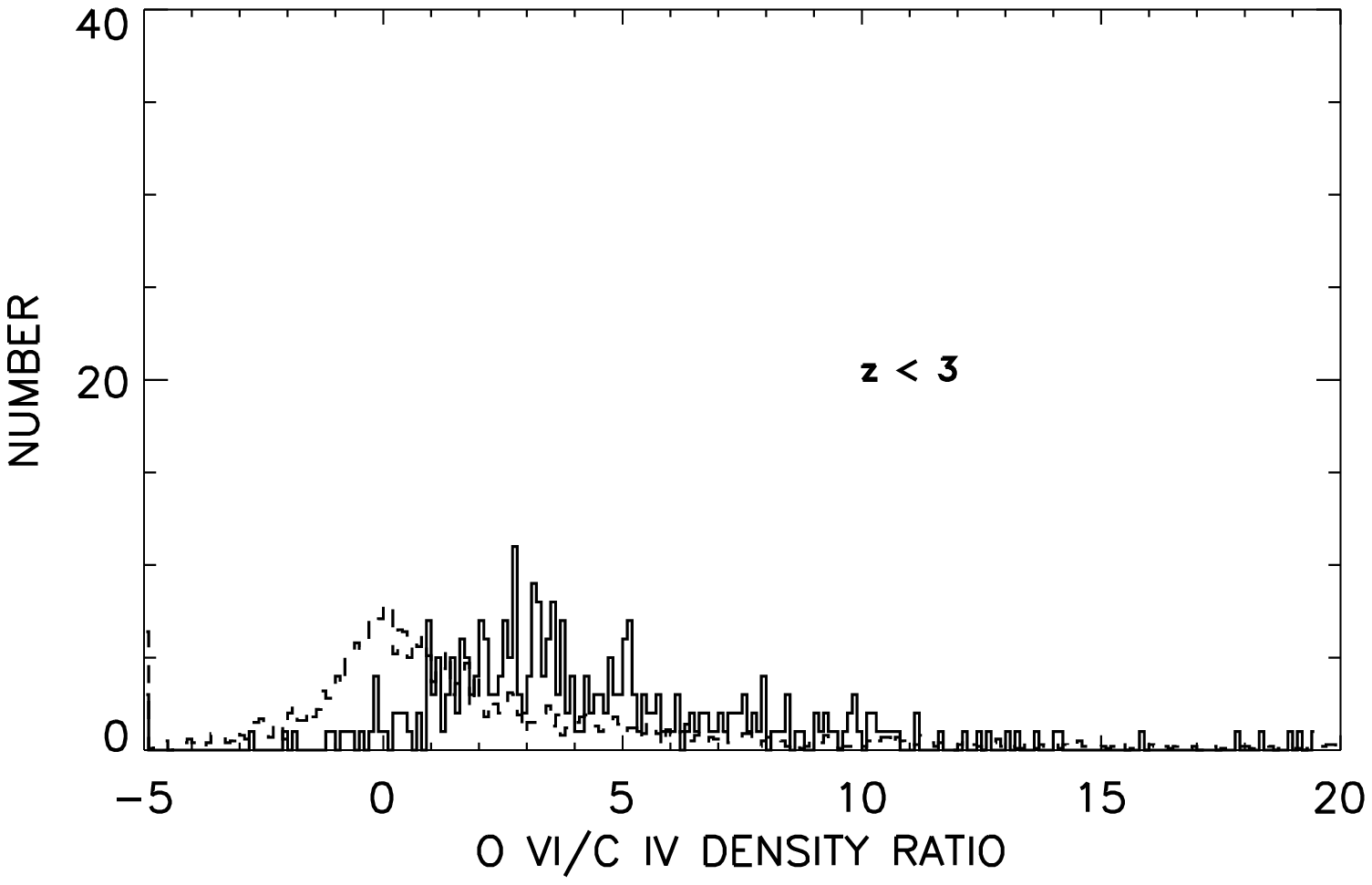}{5.5in}{0.0}{120}{120}{-300}{0}
\caption{
}
\end{figure}

\clearpage
\newpage

\begin{figure}
\figurenum{21a}
\plotfiddle{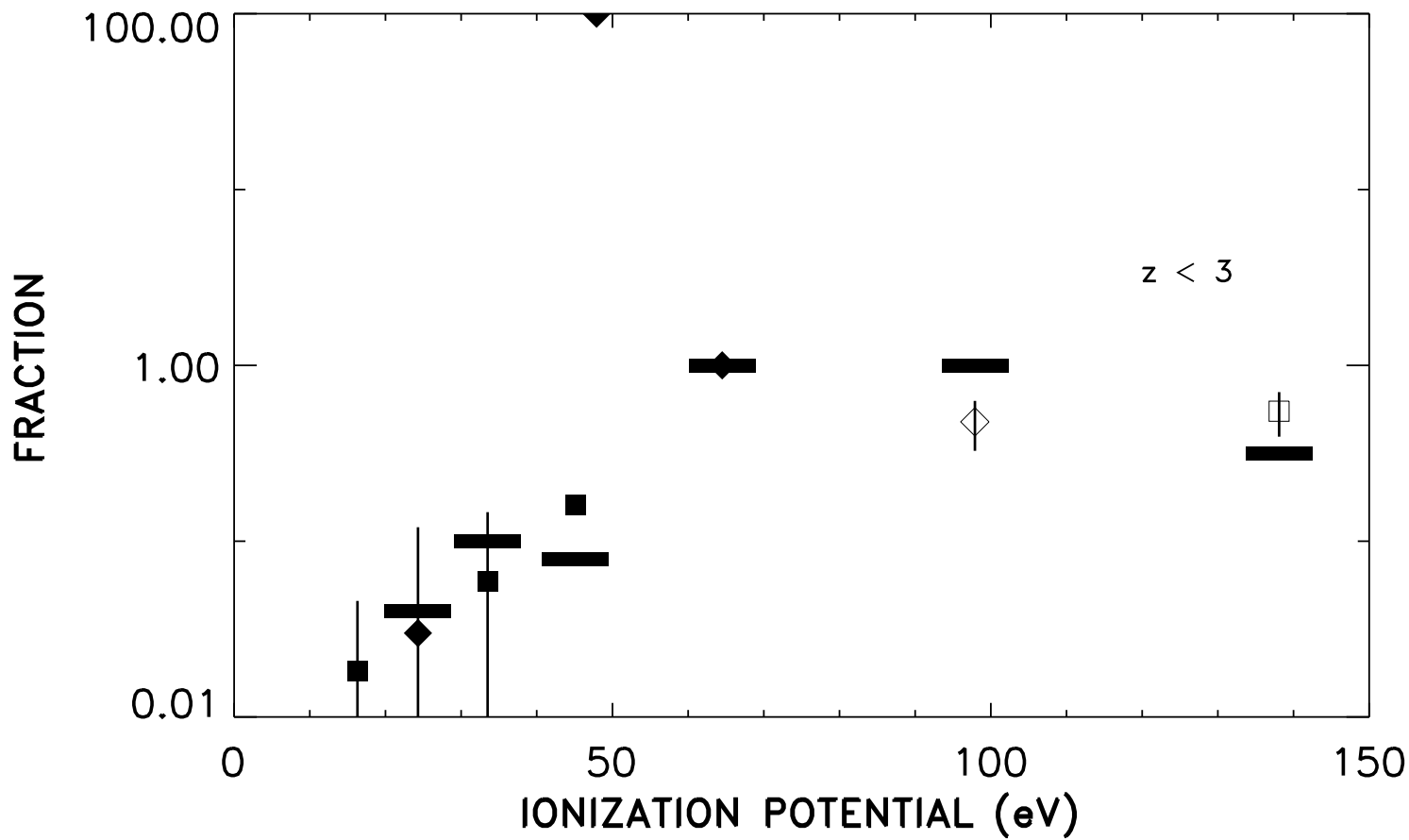}{5.5in}{0.0}{120}{120}{-300}{0}
\caption{The quantity $A_C\,N({\rm X})/A_X\,N({\rm
C~IV})$\ for $z < 3$\ for the ions, X, of Table~3, where $(A_X/A_C)$\ is the
abundance of the element $X$\ with respect to C, as a function of the
ionization potential of the ion, in eV.  Errors are $\pm 1~\sigma$.  The solid
bars show a model in which the ionizing spectrum is a $-1.8$\ power law with
$\Gamma = -1.6$.
}\label{fig:17a}
\end{figure}

\begin{figure}
\figurenum{21b}
\plotfiddle{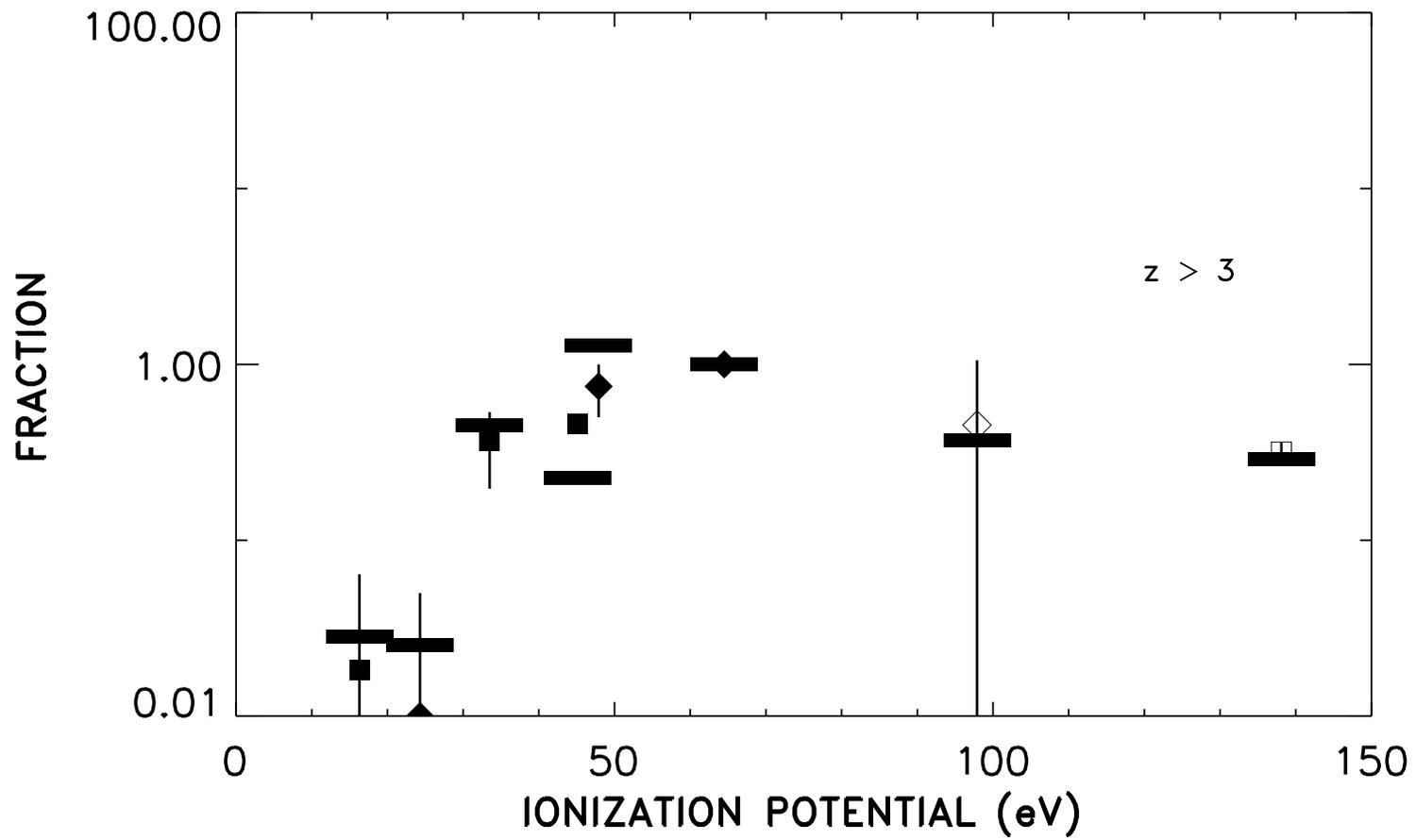}{5.5in}{0.0}{120}{120}{-300}{0}
\caption{As in (a) for $z > 3$.  See text (\S 4.3) for a
description of the comparison model (solid bars).
}\label{fig:17b}
\end{figure}

\end{landscape}

\end{document}